%% file: ms.tex
\documentclass[12pt, twocolappendix, appendixfloats, numberedappendix]{emulateapj}

\input{definition.tex}

\usepackage{natbib}
\usepackage{rotating}
\usepackage{amsmath}
\usepackage{hyperref}

\begin{document}

\shorttitle{The JHU-SDSS \mgii\ Catalog}
\shortauthors{Zhu \& M{\'e}nard}
\title {The JHU-SDSS metal absorption line catalog: \\
redshift evolution and properties of \mgii\ absorbers}

\author{
Guangtun Zhu\altaffilmark{1}
\& Brice M{\'e}nard\altaffilmark{1,2,3}
} 
\altaffiltext{1}{Department of Physics \& Astronomy, Johns Hopkins University, 3400 N. Charles Street, Baltimore, MD 21218, USA.\\ Contact: gz323@pha.jhu.edu}
\altaffiltext{2}{Institute for the Physics and Mathematics of the University of Tokyo, Kashiwa 277-8583, Japan}
\altaffiltext{3}{Alfred P. Sloan Fellow}
 
\begin{abstract}
We present a generic and fully-automatic method aimed at detecting absorption  lines in the spectra of astronomical objects. The algorithm estimates the source continuum flux using a dimensionality reduction technique, nonnegative matrix factorization, and then detects and identifies metal absorption lines. We apply it to a sample of $\sim 10^5$ quasar spectra from the Sloan Digital Sky Survey and compile a sample of $\sim 40,000$ \mgii\ \& \feii\ absorber systems, spanning the redshift range $0.4< z < 2.3$. The corresponding catalog is publicly available. We study the statistical properties of these absorber systems and find that the rest equivalent width distribution of strong \mgii\ absorbers follows an exponential distribution at all redshifts, confirming previous studies. Combining our results with recent near-infrared observations of \mgii\ absorbers we introduce a new parametrization that fully describes the incidence rate of these systems up to $z\sim5$. We find the redshift evolution of strong \mgii\ absorbers to be remarkably similar to the cosmic star formation history over $0.4<z<5.5$ (the entire redshift range covered by observations), suggesting a physical link between these two quantities.
\end{abstract}


\section {Introduction}

Metal absorption lines detected in the spectra of distant sources provide us with a powerful tool to probe the gas content in the Universe: their detectability does not depend on redshift nor the apparent luminosity of the corresponding object. They can for example be used to shed light on gas flows around galaxies. From an observational point of view the \mgiidoublet~doublet is of particular interest: it is the strongest absorption feature detectable in the optical at intermediate redshift ($0.3\lesssim z \lesssim2.5$). 
It allows us to probe low-ionization gas present in the circum- and inter-galactic media. Numerous \mgii\ surveys have been conducted  \citep[\eg][]{weymann79a, lanzetta87a, tytler87a, sargent88a, caulet89a,  steidel92a, churchill99a, churchill00a, york06a, nestor05a, prochter06a, quider11a}.
They have shown that weak $(W_0<0.3\,{\rm \AA})$ and strong $(W_0>0.3\,{\rm \AA})$ \mgii\ absorbers have different statistical properties but the nature of the absorbing gas is still debated \citep[\eg][among others]{bergeron91a, steidel94a, norman96a, churchill00b, bouche07a, chelouche10a, chen10a, chen10b, kacprzak10a, kacprzak11a, kacprzak11b, nestor11a, menard11a, bordoloi11a}, 

The Sloan Digital Sky Survey \citep[SDSS, ][]{york00a} has provided us with a sample of more than 100,000 quasar spectra well suited for the detection of intervening absorber systems. Previous works have made used of certain data releases to detect \mgii\ absorption lines with various levels of completeness and purity \citep[][]{nestor05a, york06a, bouche06a, prochter06a, lundgren09a, quider11a}.  In the era of large sky surveys it is important to develop efficient  algorithms to automatically detect absorption-line systems in quasar spectra to take advantage of the ever-growing data. In this paper we present such an algorithm and applying it to the seventh Data Release of the SDSS \citep[DR7, ][]{abazajian09a} we present the detection of $\sim40,000$ \mgii\ absorbers at $0.4<z<2.3$. With this new dataset, we study the absorber incidence rate as a function of redshift and rest equivalent width. The method presented in this paper is generic and can be applied to any large sample of spectra.

 The paper proceeds as follows: in Section \ref{sec:pipeline}, we describe the dataset  and the algorithm. We present the catalog of detected \mgii\ absorbers in Section \ref{sec:catalog} and their statistical properties in Section \ref{sec:stats}. We summarize our results in Section \ref{sec:summary}.

\section{The Absorption-line Detection Algorithm}\label{sec:pipeline}

Detecting absorption lines in the spectrum of a source requires two essential steps: (i) estimating the continuum intrinsic to the source, and (ii) detecting departures from the continuum estimate. Here we describe an algorithm performing those tasks and apply it to the SDSS quasar catalog \citep[][]{schneider10a}. This catalog includes $105,783$ spectroscopically confirmed quasars. 
We use the quasar redshift estimates provided by \citet[][]{hewett10a}\footnote{\tt http://das.sdss.org/va/Hewett\_Wild\_dr7qso\_newz/}. Besides the quasars in the DR7 catalog, \citet{hewett10a} also includes $1411$ additional visually-inspected quasars, which we treat in the same manner below. 

\subsection{Continuum estimation}\label{sec:confit}

\subsubsection{NMF eigenspectra}\label{sec:nmffit}

Studies using principle component analysis \citep[PCA, \eg][]{connolly95a} have shown that quasar spectra reside in a low-dimensional subspace \citep[\eg][]{yip04a, wild06a}. The continuum of a given quasar can be described by a linear combination of a ``small''-size basis set of eigenspectra. In order to define such a set, we use the technique of nonnegative matrix factorization \citep[NMF, ][]{lee99a, blanton07a}. Given a set of spectra, NMF defines a basis set of {\it nonnegative} eigenspectra. 
This approach is motivated by the nonnegativity of the components representing an observed quasar spectrum: continuum, emission lines and the flux of the host galaxy. In this work we choose to limit the dimensionality of the eigenspectra to twelve.
We find this value to be sufficient to capture most of the variation in the shapes of SDSS quasar spectra.
We note that this value can be increased or decreased by several without significantly changing the results of our analysis. Working with $N_{\rm dim}\ll10$ is not sufficient to capture all the complexity of quasar spectra and a very high number of dimensions sometimes provides enough flexibility to include intervening absorption lines in the source continuum estimate.

\begin{figure*}
\epsscale{1.2}
\plotone{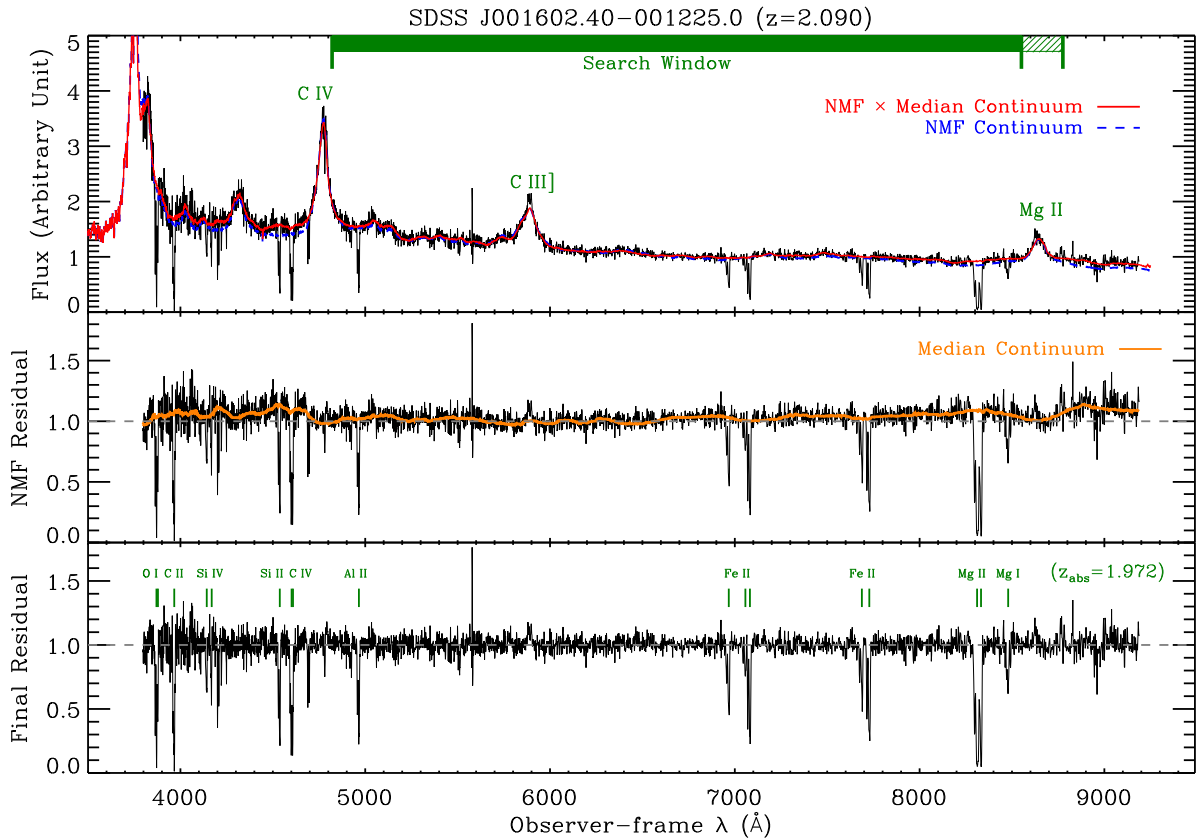}
\caption{Illustration of the steps involved in the absorption-line detection pipeline: \emph{Top panel}: The black line shows the  normalized observed spectrum for quasar SDSS J$001602.40-001225.0$. The blue line represents the best-fit NMF continuum and the red line shows the final continuum estimate after median filtering. The search window for \mgii\ absorbers is indicated at the top. The filled green region shows the region where an absorber is considered as intervening.
\emph{Middle panel}: the black line shows the NMF residual spectrum, i.e., the ratio of the observed spectrum to the NMF continuum estimate. The orange line shows the median continuum estimate. \emph{Bottom panel}: Final residual spectrum used for the absorption line detection. In this case an absorber at $z\simeq1.972$ is detected from a series of metal absorption lines.}
\label{fig:example}
\end{figure*}

Using the sample of quasar spectra introduced above we construct a basis set of NMF eigenspectra using rest-frame flux-normalized spectra. As we access different rest-frame wavelength ranges as a function of quasar redshift we cannot apply a uniform normalization. We have chosen four wavelength ranges in which quasar spectra are relatively featureless and which are enough to characterize the whole range of quasar redshifts in our sample. These regions as well as a description of our normalization scheme are presented in Appendix A. For each of these four redshift ranges we create a basis set of eigenspectra using all corresponding quasars. When estimating the continuum of a given quasar, we choose the set of eigenspectra whose median redshift is closest to the quasar's redshift. This guarantees that each quasar spectrum is described by a set of eigenspectra built from the maximal number of available quasars covering the same wavelengths. 

The presence of strongly dust-reddened quasars, broad absorption lines (BALs) or various spectroscopic artifacts can affect the eigenspectra estimation. In order to account for such outliers, we take an iterative approach. After having decomposed the spectra of all quasars into eigenvectors we keep only those for which the eigenvalues lie within $5\sigma$ of the mean eigenvalues of all input quasars. We iterate this process until no outlier is found. The code usually converges after $\lesssim10$ iterations and ensures that the construction of the eigenspectra does not include peculiar objects.

\begin{figure*}
\epsscale{1.2}
\plotone{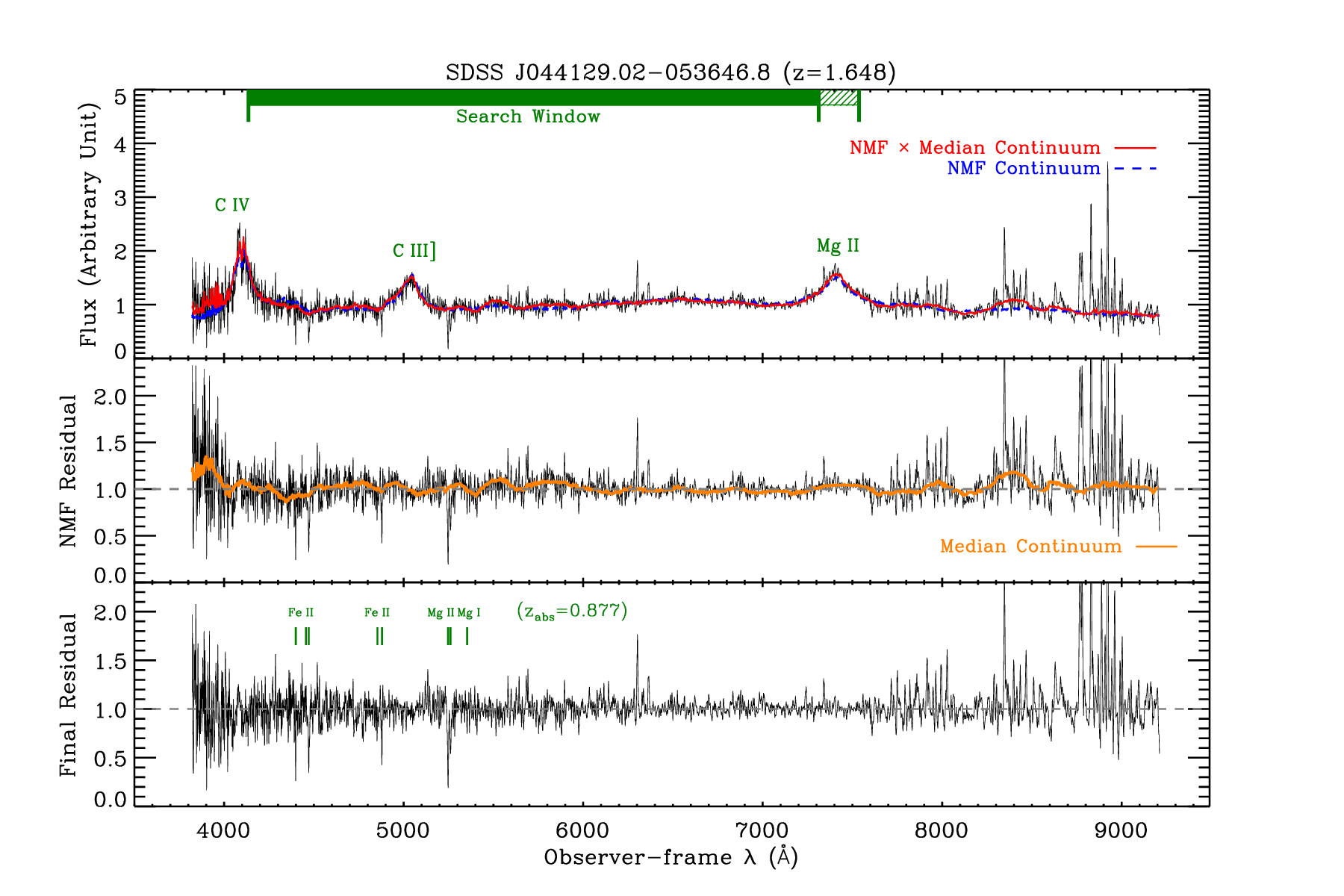}
\caption{
Same as Figure \ref{fig:example}, but for an observation severely affected by bad removal of sky emission lines (SDSS J$044129.02-053646.8$).
The residual \oi~$\lambda6300$ and OH lines in the red are conspicuous in the observed spectrum (top panel). Our continuum fitting is however not strongly affected by the presence of these features.}
\label{fig:examplesky}
\end{figure*}

Once the basis set of eigenspectra has been defined in each of the four redshift intervals, we estimate the continuum of all quasars by finding the best-fit nonnegative linear combination of the eigenspectra. We present two examples of the NMF fitting in the top panels of Figure \ref{fig:example} and \ref{fig:examplesky}.

\subsubsection{Filtering out intermediate-scale fluctuations}\label{sec:medfit}

The NMF continuum estimation captures mostly the large-scale fluctuations of a quasar spectrum. As we are interested in detecting narrow absorption lines we can improve our continuum estimation by removing power on intermediate scales. To achieve this, we apply a median filter with a size larger than the typical absorption width. 
The instrumental resolution of SDSS spectra is about $69~\kms$. The wavelength binning of the spectra is in logarithmic (velocity) space, with one pixel matching the spectral resolution.
The two \mgiidoublet~lines are separated by $7.28$ \AA~in rest frame, which translates to $\sim11$ pixels in the observer frame.  The full width at half maximum (FWHM)  of each line, convolved with the SDSS instrumental resolution,  can reach up to $500~\kms$ ($\sim7$ pixels). The overall coverage of a \mgii\ doublet is thus about $18$ pixels in the SDSS spectra. We first apply an intermediate-scale median filter with a size of $141$ pixels (about eight times the size of a strong \mgii\ absorber system) then remove smaller scales power by applying a filter with a size of $71$ pixels. While doing so we mask out pixels possibly containing narrow absorption lines by only keeping fluctuations within $1.5\sigma$ of the continuum. We repeat these two steps three times, which we found is sufficient for the estimation of the median continuum to converge.

In the middle panels of Figure \ref{fig:example} and \ref{fig:examplesky},  we show the median-filtered NMF residuals. The median filtering captures the fluctuations on intermediate scales while preserving the narrow absorption lines. In the bottom panels we show the final residuals, i.e., the spectra  normalized by the products of the NMF continua and the median continua. We also label the prominent narrow absorption lines based on the \mgii\ absorbers we find in these two spectra.  We describe the line detection method in Section \ref{sec:linedetect}.

\begin{figure*}
\epsscale{1.}
\plotone{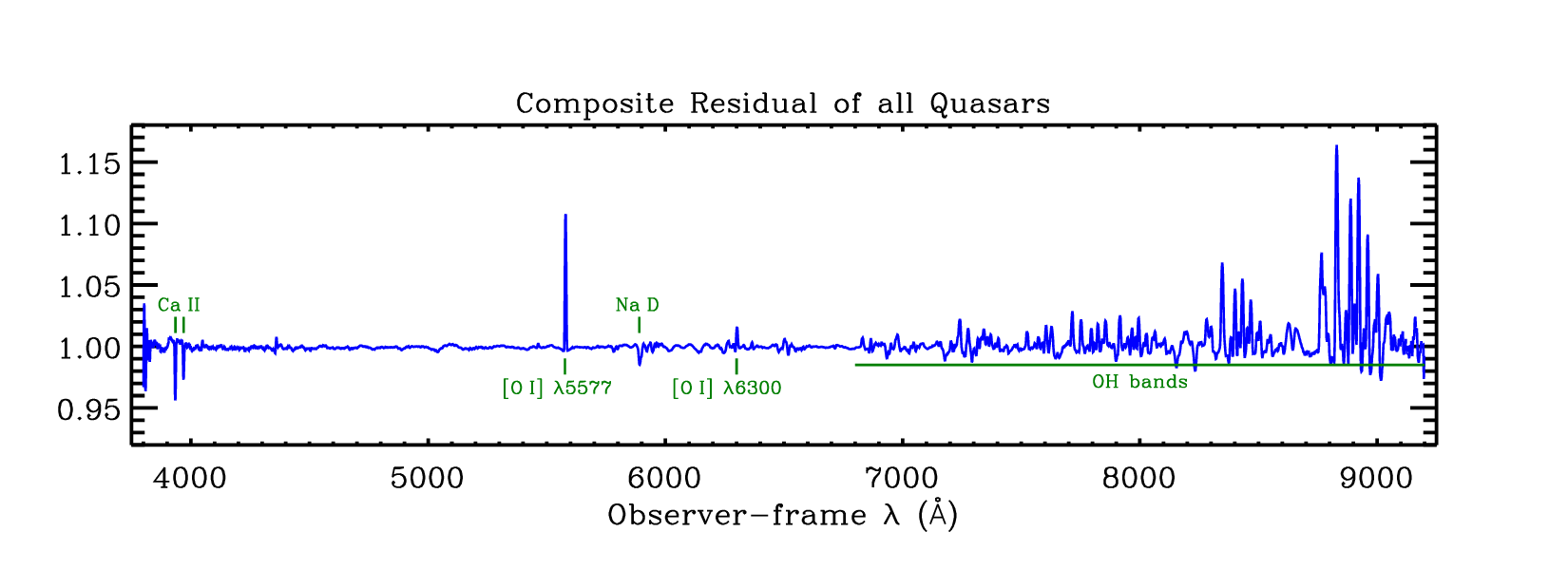}
\caption{The composite residual spectrum of all quasars in the observer frame. This shows that the spectral extraction of SDSS on average underestimated sky emission lines, \eg \oi~$\lambda5577$, \oi~$\lambda6300$, and OH lines. The absorption lines \caii~and Na D are caused mainly by interstellar medium in the Milky Way. 
}
\label{fig:residual}
\end{figure*}

\subsubsection{Sky emission and galactic absorption}\label{sec:skyresidual}

SDSS spectra contain features due to sky emission lines such as the \oi~$\lambda5577$ and OH lines that are not properly subtracted. This can be seen in Figure \ref{fig:examplesky}. These features do not substantially affect our estimation of the quasar continua which is done in the quasar rest frame. However, care is needed when detecting narrow absorption lines and estimating their completeness.

To quantify this effect we stack all the residual spectra in the observer frame. The result is shown in Figure \ref{fig:residual}.  It shows that the SDSS pipeline on average under-subtracts sky emission lines, \oi~$\lambda5577$, \oi~$\lambda6300$,  and OH lines\footnote{See also \citealt{yan11a}. This effect is known to the SDSS reduction team and the noise estimate of the flux is  enhanced accordingly (David Schlegel, private communication).}. In addition, the \caiidoublet~and Na D
absorption lines induced by the interstellar medium (ISM) of the Milky Way are clearly visible.  When searching for narrow absorption lines we need to exclude  the \caii~region that may introduce false positives. When determining the completeness of our pipeline,  we also need to account for all these potential sky residuals. We will come back to this in Section \ref{sec:completeness}.

\subsubsection{Outlier rejection}\label{sec:rejection}

A fraction of quasar spectra contain BALs which complicate the continuum estimation. To identify them we measure the variance of each quasar flux residual and exclude objects for which the value is significantly larger than that of the overall population. This procedure removes $16,704$ objects from our sample. We note that this in principle can reject quasars whose spectra host a  very large number ($\sim10$) of strong \mgii\ absorbers.  These systems, however, are extremely rare, and should not have any practical  effects on our survey. 
Due to catastrophic errors and gaps in the data, a small fraction of the quasar spectra ($55$ objects) present less than 5 valid pixels in the wavelength ranges used for flux normalization. Such objects are not included in our analysis. 
Beyond $z=4.7$, we have only $219$ quasars and cannot build a well-defined basis set of eigenspectra. We do not consider these high-redshift quasars. We also exclude $5682$ quasars with $z<0.4$ which cannot be used to look for \mgii\ absorption. This leaves $84,534$ quasars well suited for narrow absorption-line detection.

\subsection{Absorption line detection}\label{sec:linedetect}

Having compiled a set of continuum-normalized quasar fluxes we now detect, identify and characterize narrow absorption lines. Our procedure includes three steps:  (1) candidate selection; (2) false positive elimination;  and (3) equivalent-width measurement.

\subsubsection{Search window}\label{sec:searchwindow}

For a given quasar spectrum the redshift range in which we search for absorbers is constrained by several factors: the wavelength coverage of the SDSS spectrum, the redshift of the quasar, and the capability of the detection method to differentiate between different types of absorbers. 

\mgii\ absorbers with $z_{\rm abs}\sim z_{\rm QSO}$  are  likely physically associated with their background quasar. Associated  absorbers can either be blueshifted or redshifted \citep[\eg][]{vandenberk08a, shen12a}. Although we are primarily interested in intervening \mgii\ absorbers  associated with foreground sources, we extend the search window redshifted from the quasar by $\Delta z=0.04$ ($12,000~\kms$) to include these  quasar-associated absorbers.  At wavelengths blueward of the quasar \civ~emission line, the covering fraction of  intervening \civ~absorbers is substantially higher than that of \mgii\ absorbers. A doublet found close to or blueward of the \civ~emission line, has a higher probability to be \civ~ than \mgii. In this \mgii-based survey, we thus do not consider the region blueward  of the quasar's \civ~line, leaving  the \civ-\mgii\ discrimination to future work.  Since \civ~absorption lines can also be redshifted, we  conservatively start the search window redward of \civ~by $\Delta z=0.02$ ($6000~\kms$).

The ISM in the Milky Way can also cause absorption lines in the spectra of extragalactic sources (Figure \ref{fig:residual}). Our experience shows that in some cases the \caiidoublet~lines from the Milky Way can mimic a $z\sim0.4$ \mgii\ doublet. To avoid the introduction of such false positives, we mask out the \caii~region in our search window for \mgii\ absorbers.

Our final search window in each quasar spectrum therefore starts  from $\Delta z=0.02$ redshifted from the quasar's \civ~emission line  or the blue end of the SDSS coverage ($\sim3800$ \AA), and ends at  $\Delta z=0.04$ redshifted from the quasar's \mgii\ emission line or the red end of the SDSS coverage ($\sim9200$ \AA), excluding the  observer-frame \caii~regions. In the top panels in Figure \ref{fig:example} and \ref{fig:examplesky},  we show the search windows of the given examples.

\subsubsection{First pass: candidate selection}\label{sec:linecriteria}

The first step in the line detection is to select a list of absorption line candidates. To do so we use a multi-line model including \mgiidoublet~and four strong \feii~lines: $\lambda2344$, $\lambda2383$, $\lambda2586$,  and $\lambda2600$. The inclusion of these \feii~lines facilitates the elimination of false positives (see next sub-section).
We then perform a match filter search for candidates detected above a certain signal-to-noise ($SNR$ ) ratio threshold. Within the search window of a given quasar, we convolve the residuals and the noise estimates with the multi-line model using top-hat filters with a width of $4$ pixels ($276~\kms$) for each line.  
This is motivated by the typical FWHM of the absorption lines which, convolved with the SDSS instrumental resolution, is $\sim100-400~\kms$  ($\sim2-6$ pixels). For each quasar, we perform the convolution at every potential absorber redshift, given by all the pixels within the search window.  We then select absorber candidates at pixels that satisfy 
${\mathrm  \criterionmgii}:$
\begin{eqnarray}
{SNR}(\mathrm{Mg~II}~\lambda2796) > 4~~\&\&~~SNR(\mathrm{Mg~II}~\lambda2803) > 2 {\mathrm .}\nonumber \\
\label{eq:criterionmgii}
\end{eqnarray}

\noindent This criterion determines the window function of our search. Finally, for candidates with continuous  redshifts/pixels, we group them together and treat them as one single  candidate with their median redshift.

\subsubsection{Second pass: false positive elimination}\label{sec:lineelimination}

Once we have a list of absorber candidates that passed \criterionmgii, we take the following steps to eliminate false positives: 
(i) We fit each \mgiidoublet~doublet candidate with a double-Gaussian profile and reject candidates with peculiar separations between two Gaussians. In the fitting, we assume the dispersions of the two Gaussians to be the same but allow their centers and amplitudes to be different.  We reject a candidate if the separation between the two Gaussians differs from the fiducial value by $1$ \AA. Experiments show that the exact value of this criterion has little effect and this method efficiently eliminates the majority of false positives.
(ii) To strengthen the identification of a \mgii\ absorber 
we make use of the \feii~lines.
For each quasar, we compare every two remaining candidates and examine if any of the \feii~lines from one candidate is at the same wavelength as any of the \mgii\ lines from the other. If so, we rank the two candidates by the average $SNR$ of  the four absorption lines: the \mgiidoublet, \feii~$\lambda2600$, and $\lambda2586$. We then keep the one with the higher average $SNR$ {\it unless} the other  one has all the four \feii~lines ($\lambda2344$, $\lambda2383$, $\lambda2586$, and $\lambda2600$) detected above $2\sigma$, i.e., unless it satisfies 
${\mathrm  \criterionfeii:}$
\begin{eqnarray}
{SNR}(\mathrm{Fe~II}~\lambda2344, ~\lambda2383, ~\lambda2586, ~\& ~\lambda2600) > 2 \mathrm{,}~~~ 
\label{eq:criterionfeii}
\end{eqnarray}

\noindent in which case we keep both. Since a false positive caused by a line confusion does not have  other lines at the right wavelengths, it has a lower average $SNR$ of  the four lines and this method therefore efficiently eliminates the remaining line confusions.
(iii) In some rare cases the \ciii\ emission line of some quasars is not properly modeled by our procedure and gives rise to absorption-like features in the residuals. To eliminate these false positives, besides \criterionmgii~we require an additional detection of one of the four \feii~lines above $2\sigma$ if a candidate is at redshifts $\Delta z=\pm0.02$ from its host's \ciii. This additional criterion decreases the completeness of \mgii\  absorption-systems with weak \feii~lines in \ciii~regions. When evaluating the completeness of our survey (Section \ref{sec:completeness}), we will exclude this region.

\subsubsection{Final pass: absorber properties}
\label{sec:linemeasurement}

Having a list of robust \mgii\ absorber systems, we now determine their redshift and line properties. We estimate the rest equivalent width of each available absorption line by fitting a Gaussian profile. When two line profiles overlap we perform the fit with a double-Gaussian profile to prevent biases in the rest equivalent width estimation. This procedure also allows us to estimate the redshift and rest equivalent widths of each \mgii\ system more precisely than done in the first pass.

\begin{figure*}[!t]
\epsscale{0.43}
\plotone{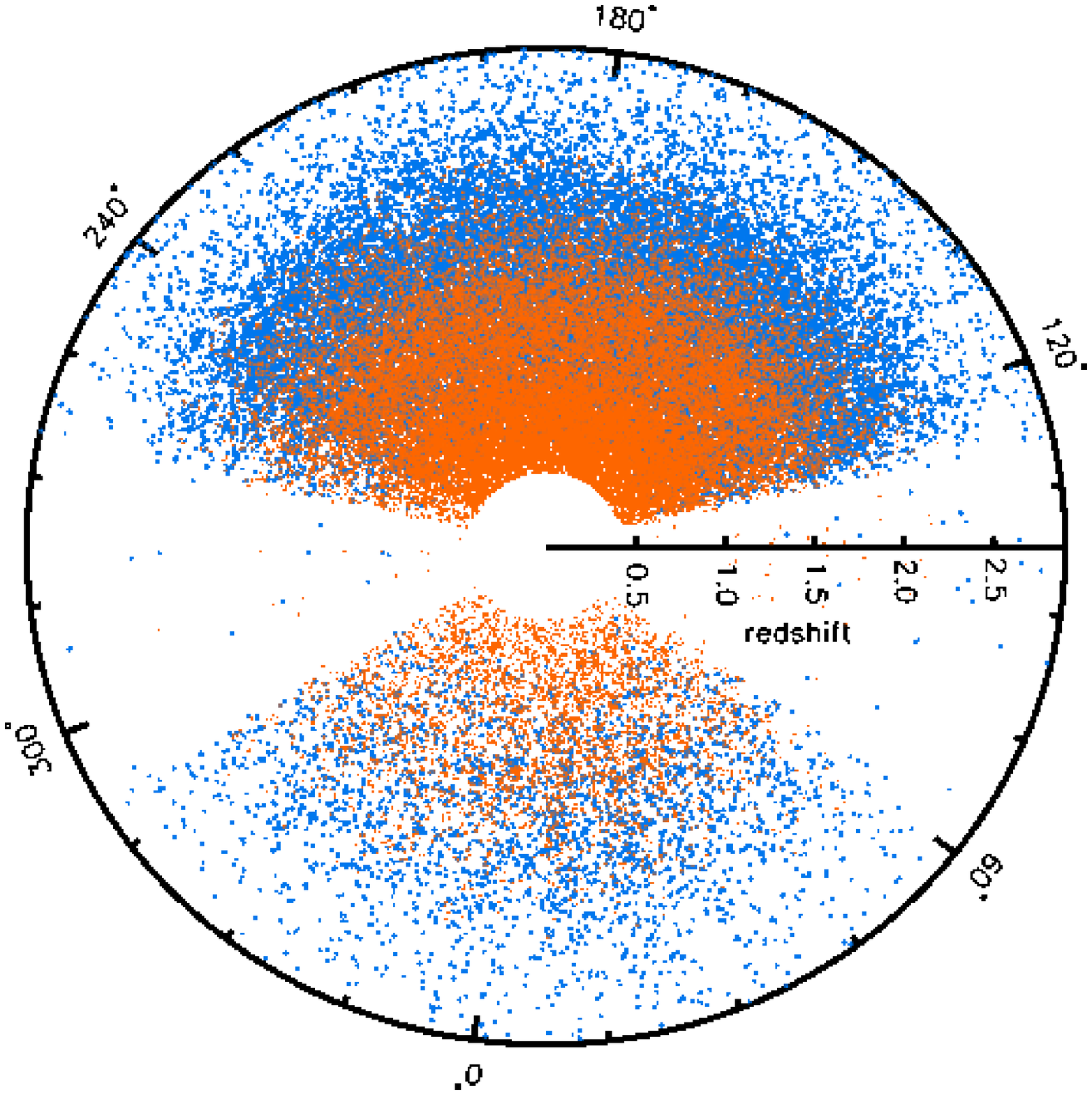}
\hspace{1.5cm}
\epsscale{0.55}
\plotone{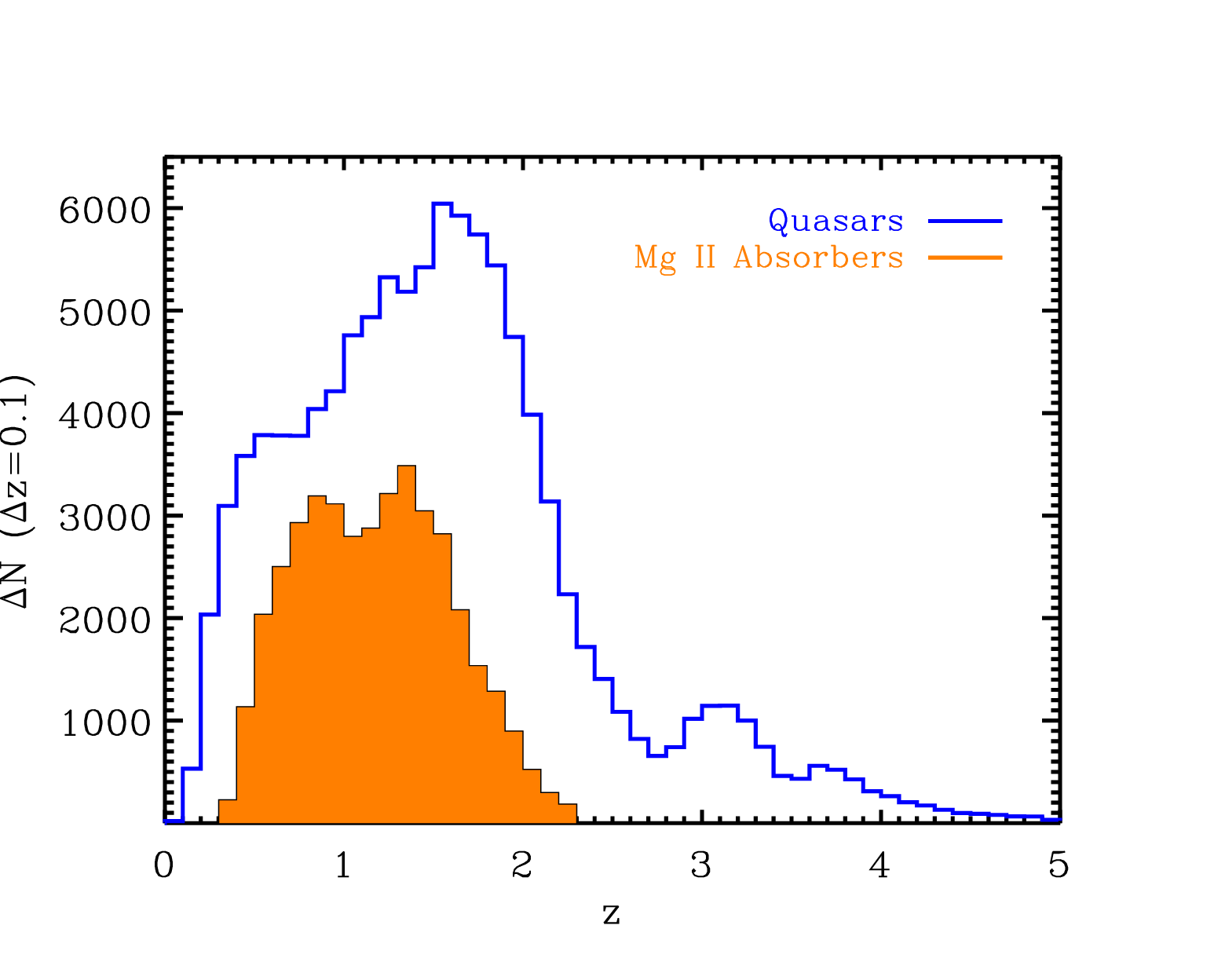}
\caption{\emph{Left}: pie diagram of the quasars (blue) and  \mgii\ absorbers (orange) in the RA-$z$ space. For clarity we only show quasars with a detected absorber. \emph{Right}: redshift distributions of all $107,194$ quasars and 
$40,429$ \mgii\ absorbers, spanning redshift from $z=0.36$ to $z=2.29$.}
\label{fig:zhist}
\end{figure*}

\section{The \mgii\ absorber sample}\label{sec:catalog}

We ran our line detection pipeline on the $84,534$ (out of $107,194$) quasars  suitable for narrow absorption line detection. Within the search window,  we detect $40,429$ \mgii\ absorbers. The spatial distribution of the quasars and absorbers are shown in the left panel of Figure~\ref{fig:zhist} with orange and blue points, respectively. The corresponding redshift distributions are shown in the right panel.

In this section, we will focus on so-called \emph{intervening} absorbers. We conservatively define such systems to be blue-shifted from their background quasar by at least $\Delta z=0.04$, which corresponds to $\Delta v\sim12,000\kms$. This absorber sample has $35,752$ objects. We now characterize its completeness and purity.

\subsection{Comparison with the Pittsburgh Catalog}\label{sec:comppitts}

\begin{figure*}[t]
\vspace{-0.5cm}
\epsscale{0.9}
\plotone{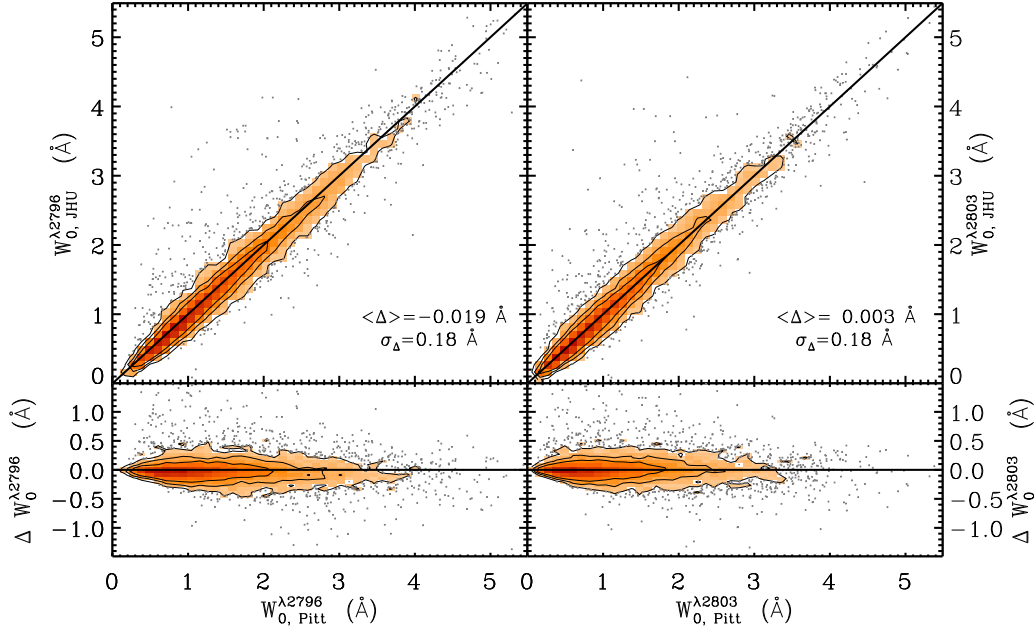}
\caption{Comparison of the rest equivalent width measurements of 
 \mgii\ $\lambda2796$ (left) and  \mgii\ $\lambda2803$ (right) between the present work and the Pittsburgh catalog.
The contours enclose $70\%$, $85\%$, and $95\%$ of the sample in each panel. The lower panels show the difference  $\Delta W_0^= W_{0, \mathrm{JHU}}- W_{0, \mathrm{Pitts}}$. The mean differences $\langle\Delta\rangle$ and sample dispersions $\sigma_\Delta$ are shown in the upper panels for clarity. 
The measurements by the two pipelines agree very well with no systematic shift and a $\lesssim 0.2$ \AA\ scatter, which is the typical measurement error.}
\label{fig:wwcomppitts}
\end{figure*}

Prior to this work the largest compilation of \mgii\ absorbers is the Pittsburgh catalog based on the SDSS DR4 dataset  \citep[][]{quider11a}\footnote{\tt http://enki.phyast.pitt.edu/PittSDSSMgIIcat.php}, using the detectionmethod presented in \citet{nestor05a}. To ensure a high purity and completeness of the absorber detection these authors visually inspected the quasar flux residuals. This sample therefore provides us with a good test bed for our pipeline.
There are $41,881$ common quasars searched  for \mgii\ doublets in both surveys. Within the search winow, we  detected $18,748$ \mgii\ absorbers with $\rewmgiione>0.02$ \AA, while the  Pittsburgh group detected $14,715$. Among these $14,715$ detections, we recovered $14,079$ ($\sim95\%$). The remaining $\sim5\%$ did not pass our \criterionmgii\ due to noise or masks. As this effect is taken into account in our completeness estimation, these missing systems do not bias any statistical analysis. In addition, we detected $4669$ systems ($\sim25\%$) that are not included in the Pittsburgh catalog but are fully consistent with \mgii\ absorbers\footnote{It is known that some absorbers were missed due to human errors and will be included in their future data release (Daniel Nestor, private communication)}. In Appendix B, we carefully inspect these systems and demonstrate that they are bonafide \mgii\ absorbers.
In Figure \ref{fig:wwcomppitts}, we compare the rest equivalent width  measurements, \rewmgiione~and \rewmgiitwo, for common absorbers in  both catalogs. The two rest equivalent width distributions appear to be consistent and the scatter of the residuals is comparable to the typical measurement error.

\begin{figure*}[h]
\epsscale{1}
\epsscale{0.5}
\plotone{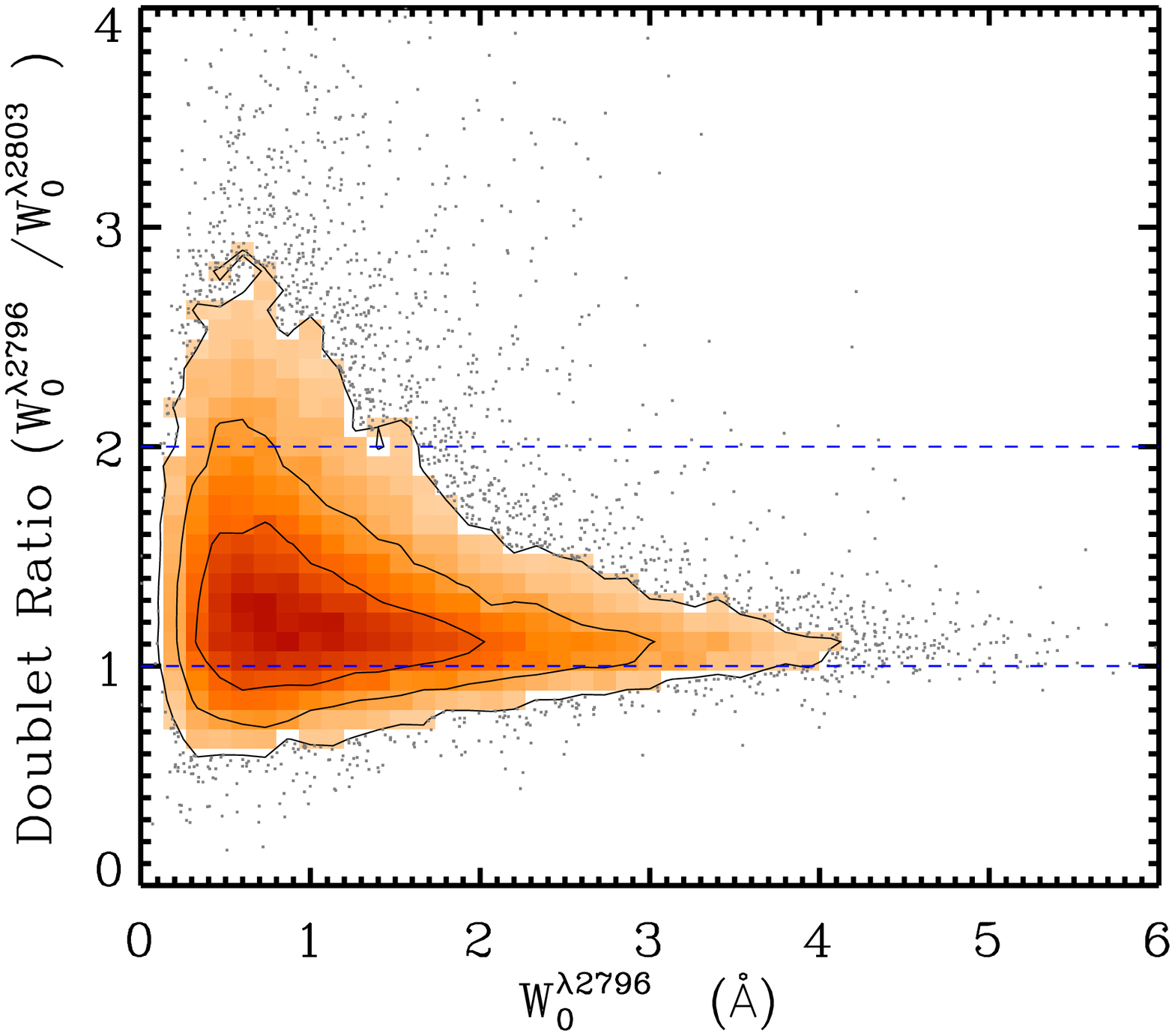}
\epsscale{0.5}
\plotone{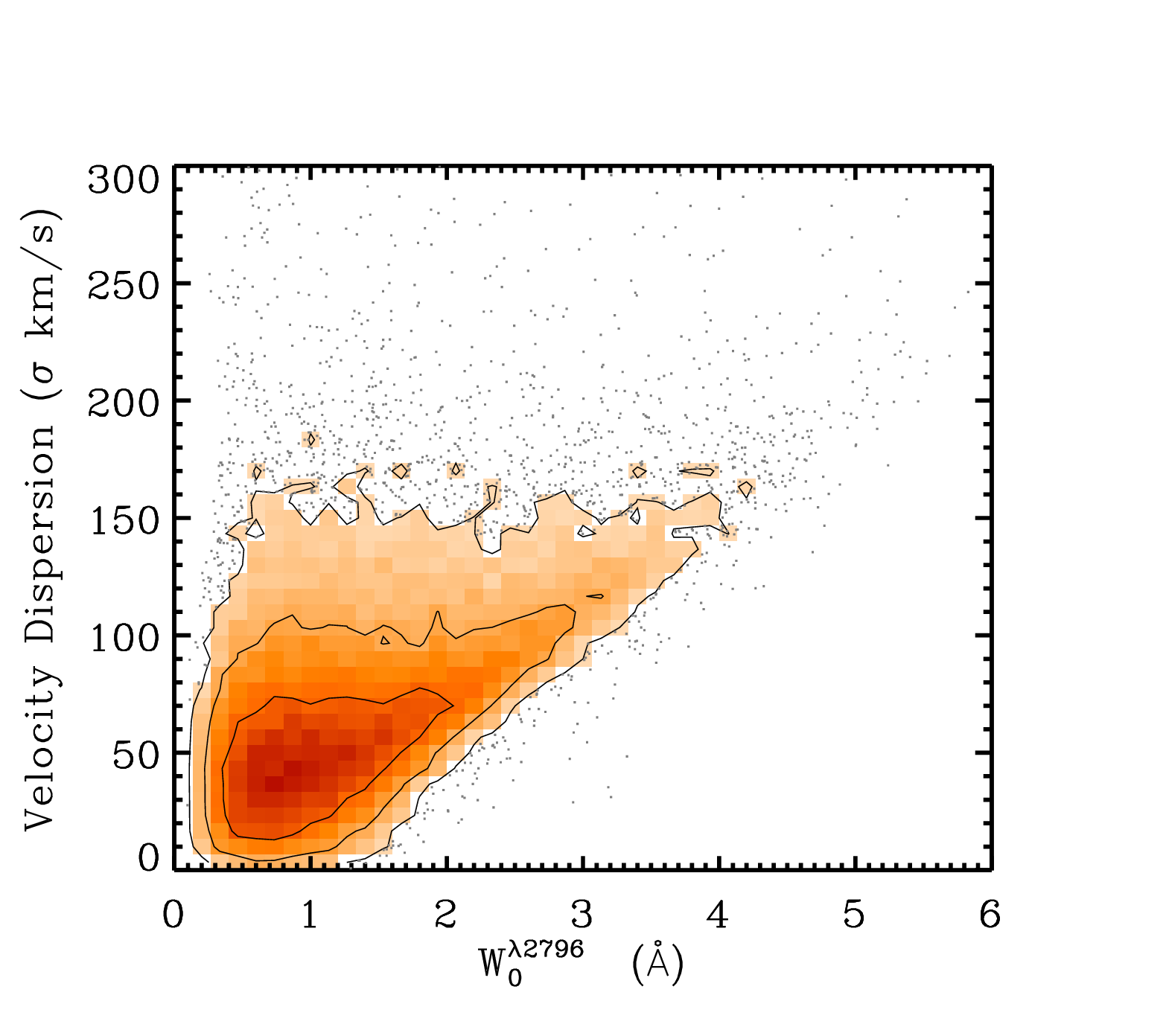}
\caption{
\emph{Left panel} :
The \mgiidoublet~doublet ratio distribution. The contours enclose $50\%$, $80\%$, and $95\%$ of the sample. The two blue horizontal dashed lines show the two theoretical limits $1$ and $2$.  Most of the doublets are saturated with ratio $\sim 1$,  indicating the rest equivalent width primarily measures velocity spread rather than column density.  The fraction of unsaturated ones increases  towards the weaker end.
\emph{Right panel}:
Velocity dispersion distribution of the \mgiidoublet~doublets from  the double-Gaussian fitting. The velocity dispersion scales nearly linearly with \rewmgiione. In both panels the contours enclose $50\%$, $80\%$, and  $95\%$ of the sample.}
\label{fig:dratiovdisp}
\end{figure*}

\subsection{Properties}\label{sec:props}

The saturation level of the \mgii\ doublet is a valuable indicator. The $\rewmgiione/\rewmgiitwo$ ratio is expected to be bounded between 2 (optically thin regime) and 1 (saturated).
We show the doublet ratio distribution of our catalog as a  function of \rewmgiione~in the left panel of Figure \ref{fig:dratiovdisp}. For comparison, we also overplot the expected minimum and maximum values with horizontal dashed lines.  The distribution shows that  most of the doublets are saturated, especially at $\rewmgiione\gtrsim1$ \AA.   The rest equivalent widths of most of these doublets therefore measure the  kinematics of the ionized gas. The fraction of unsaturated absorbers increases towards the weaker end.

In the right panel of Figure \ref{fig:dratiovdisp},  we show the measured Gaussian velocity dispersion as a function of  \rewmgiione.  We have removed the SDSS  instrumental resolution $69~\kms$ by quadrature subtraction.   Some systems end up with a negative velocity dispersion due to the noise level. They are not shown in the figure. The Gaussian velocity dispersion  scales nearly linearly with \rewmgiione, especially at the strong end. This is expected since  equivalent width primarily measures the velocity spread of the gas.

\begin{figure*}[t]
\epsscale{0.5}
\plotone{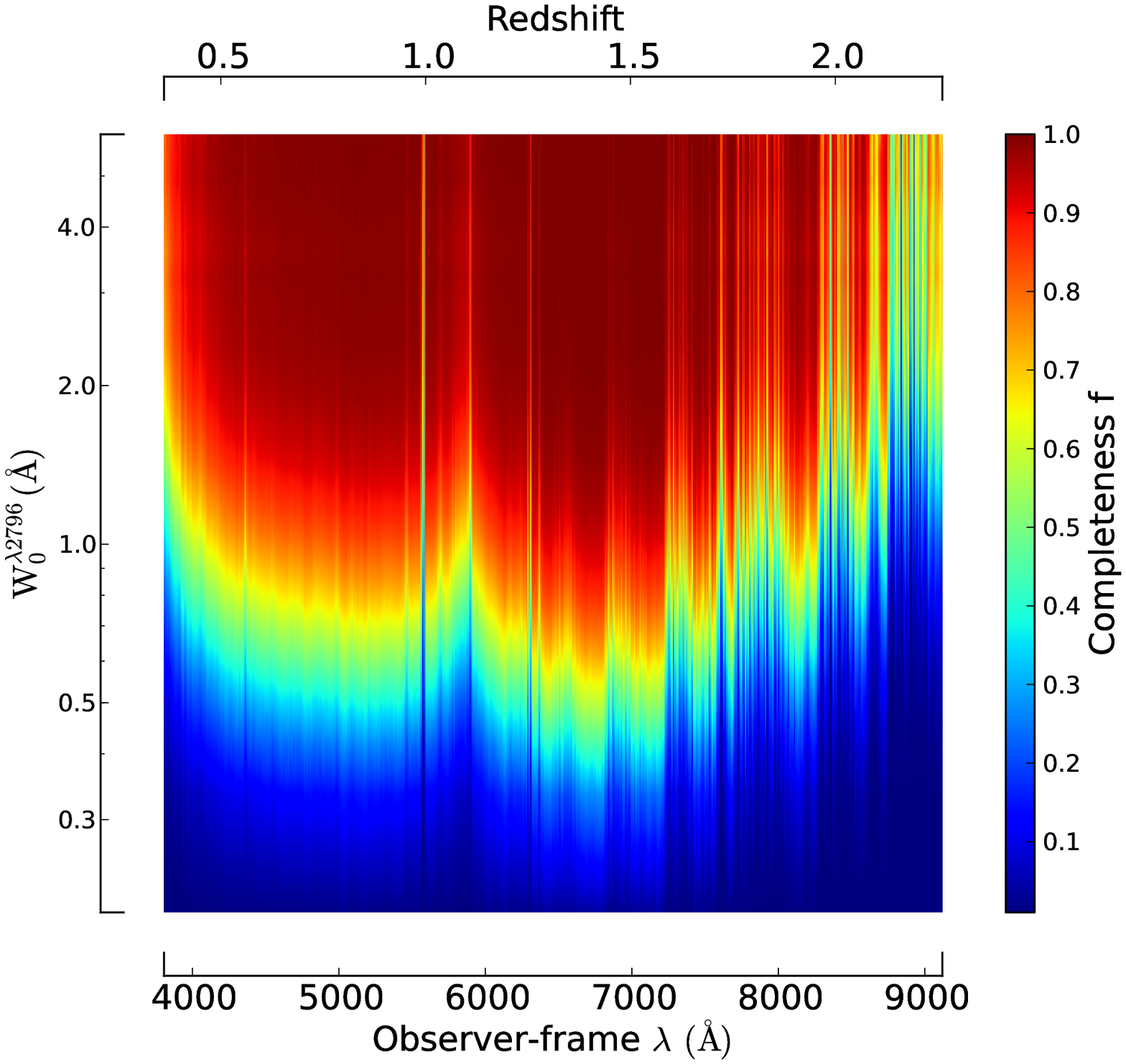}
\hspace{1.25cm}
\epsscale{0.44}
\plotone{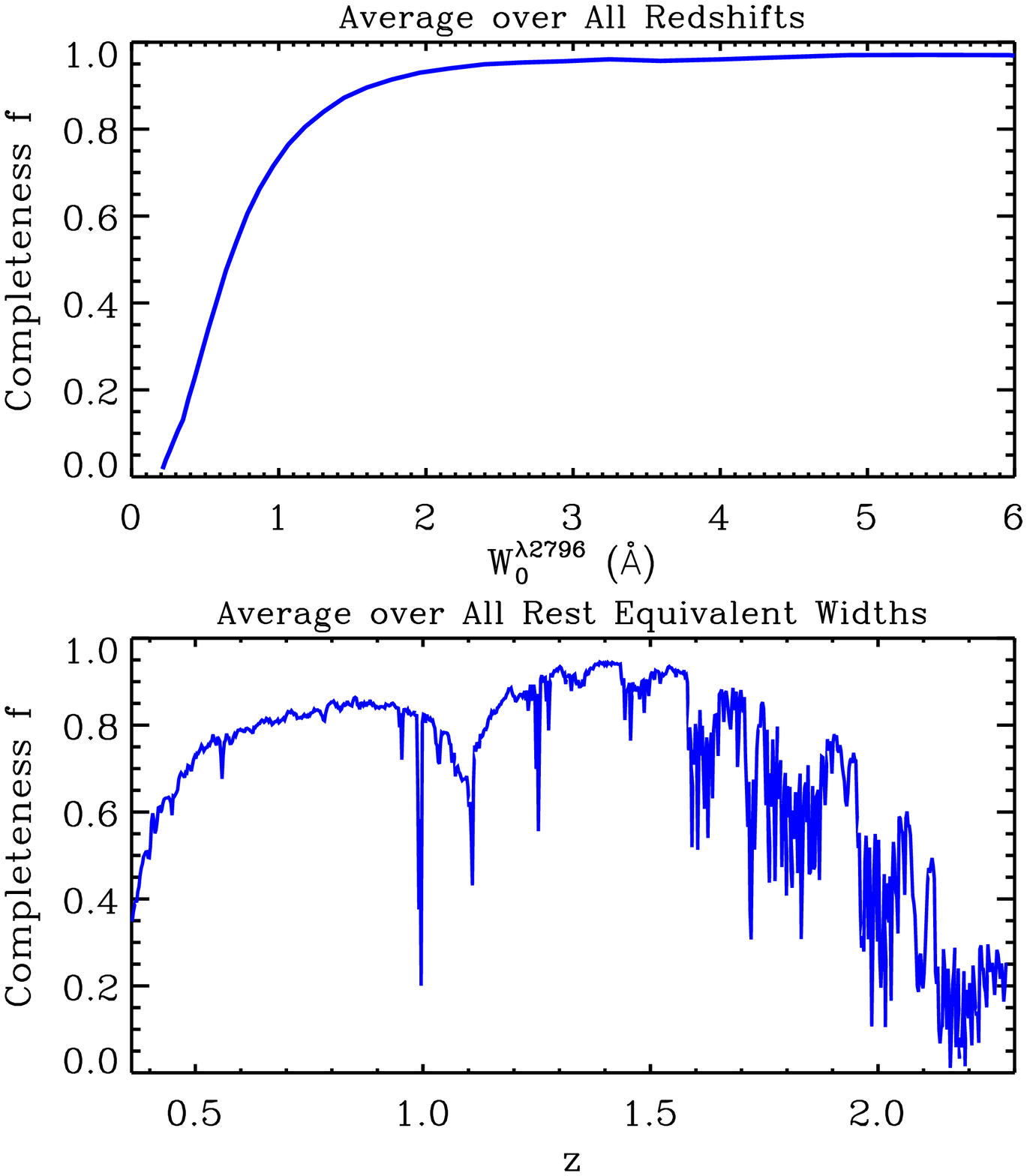}
\caption{
\emph{Left:} The completeness $f(\rewmgiione,z)$ in the \rewmgiione-$z$ space. The low completeness due to sky lines, e.g., \oi~$\lambda5577$, high pressure sodium bump at $\sim 5900$ \AA, \oi~$\lambda6300$, and  OH lines in the red, is clearly visible.  \emph{Top right:} Average completeness $f(\rewmgiione)$ over all redshifts.
\emph{Bottom right:} Average completeness $f(z)$ over all rest equivalent widths.}
\label{fig:completeness}
\end{figure*}

\subsection{Completeness}\label{sec:completeness}

The detection of an absorber with a given rest equivalent width, doublet ratio, and redshift depends on the accuracy with which the source continuum can be estimated. We now estimate the detection completeness of our algorithm using a Monte Carlo simulation.  

\begin{figure}
\epsscale{1.35}
\plotone{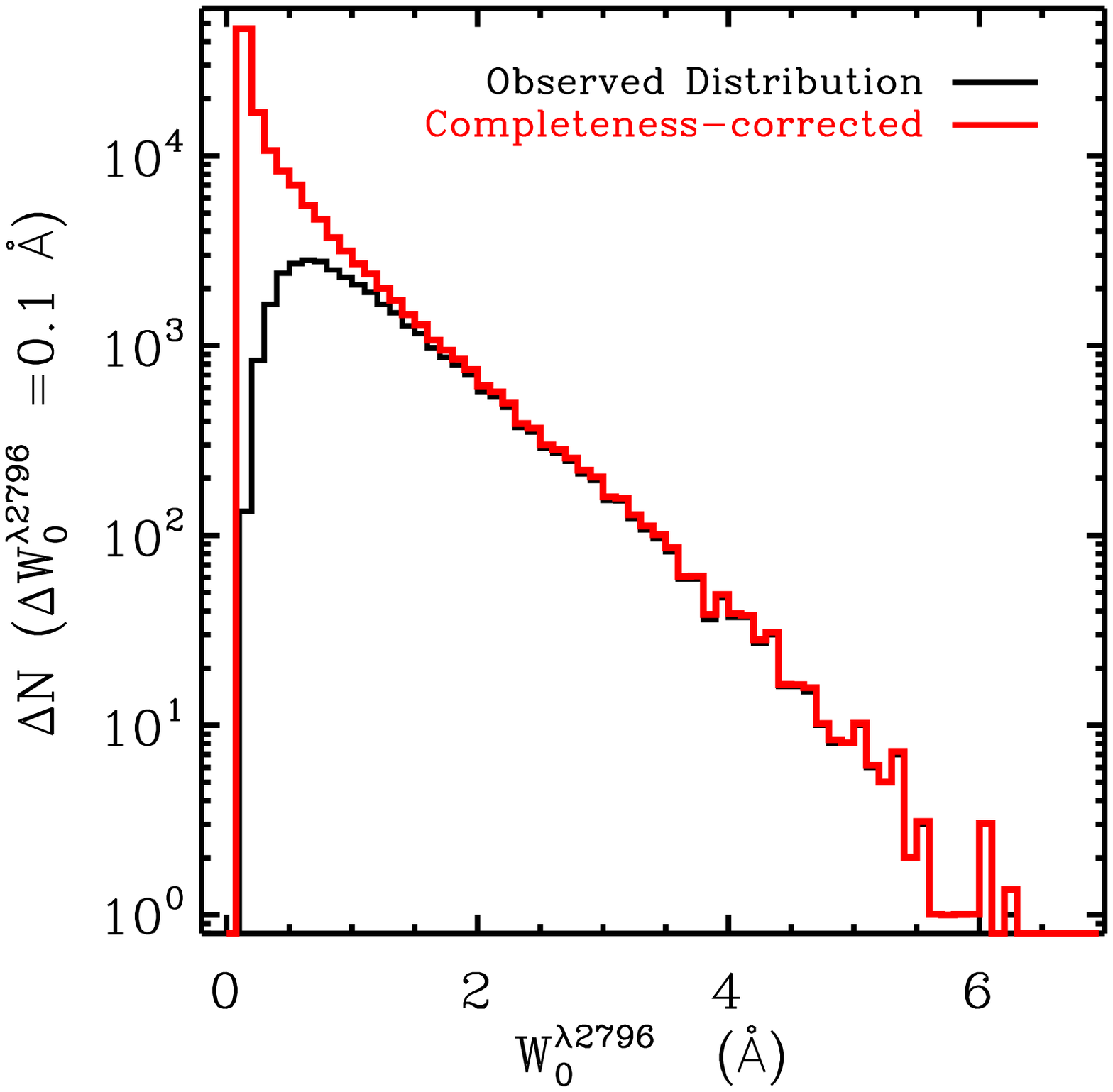}
\caption{
Distribution of \rewmgiione\ rest equivalent widths. The black line shows the observed distribution and the red line shows the intrinsic distribution 
after completeness correction.}
\label{fig:whist}
\end{figure}

We simulate absorbers drawn from a distribution of rest equivalent widths and doublet ratios. For each quasar, at each pixel, we insert a fake absorber into the flux residuals. We consider this absorber {\it covered} by the spectrum if relevant pixels  are not masked out, and {\it detected} if its final signals pass the  \criterionmgii~compared to the convolved noise model at those pixels. With the Monte Carlo simulation, we determine the average redshift path,  given a redshift bin $\Delta z$, as the bin width multiplied by the  fraction of {\it covered} absorbers: $\overline{\Delta z}=\Delta z f_{\mathrm{covered}}$. At a given rest equivalent width and redshift, we determine the completeness  $f(\rewmgiione,z)$ as the ratio of the number of {\it detected} absorbers to that of {\it covered} absorbers, marginalized over all doublet ratios.
We present the completeness $f(\rewmgiione,z)$ in the  $\rewmgiione-z$ space in the left panel of Figure \ref{fig:completeness}.   In the right panels, we present the  completeness as a function of \rewmgiione~ and $z$ averaged  over all redshifts and all rest equivalent widths in the upper and  lower panels, respectively. The  completeness is higher for stronger absorbers and at redshifts for which the noise level of the flux residuals is lower.  The conspicuous low completeness spikes in the left panel  (dips in the bottom right panel) are due to prominent sky lines, e.g., \oi~$\lambda5577$ and OH lines in the red.  The broad low completeness bump at $\sim1.0-1.2$ is   caused by a combination of high-pressure sodium at  $\sim5500-6100$ \AA~in the sky light and the decreasing sensitivity during the split of the blue and red spectrographs\footnote{\tt http://www.sdss.org/dr7/instruments/spectrographs}. Towards both ends, the sensitivity of the SDSS spectrographs drops thus reducing the completeness.
We can now derive the intrinsic incidence rate of \mgii\ absorbers from the detected absorbers by weighting each absorber with \rewmgiione~at redshift $z$ with  $w=1/f(\rewmgiione, z)$. In Figure \ref{fig:whist}, we show that the observed (black) and intrinsic (red) \rewmgiione\ incidence distributions.

\section{Statistical properties}\label{sec:stats}

The incidence rate \dndzdw\ of \mgii\ absorbers, i.e. the number of systems per unit redshift and rest equivalent width, carries important information on the number density and crosssection of the absorber systems, as a function of redshift. As pointed out by \cite{nestor05a}, the distribution of \mgii\ rest equivalent widths is well described by an exponential distribution above $W_0\gtrsim0.3{\rm \AA}$ while weaker absorbers follow a power-law distribution. The two populations may be described more generically using a Schechter function \citep[\eg][]{kacprzak11c}.  Using our absorber sample we now focus on strong \mgii\ absorbers and study their incidence rate  \dndzdw.

\subsection{Rest equivalent width distribution}
\label{sec:dndzdw}

\begin{figure}
\epsscale{1.35}
\plotone{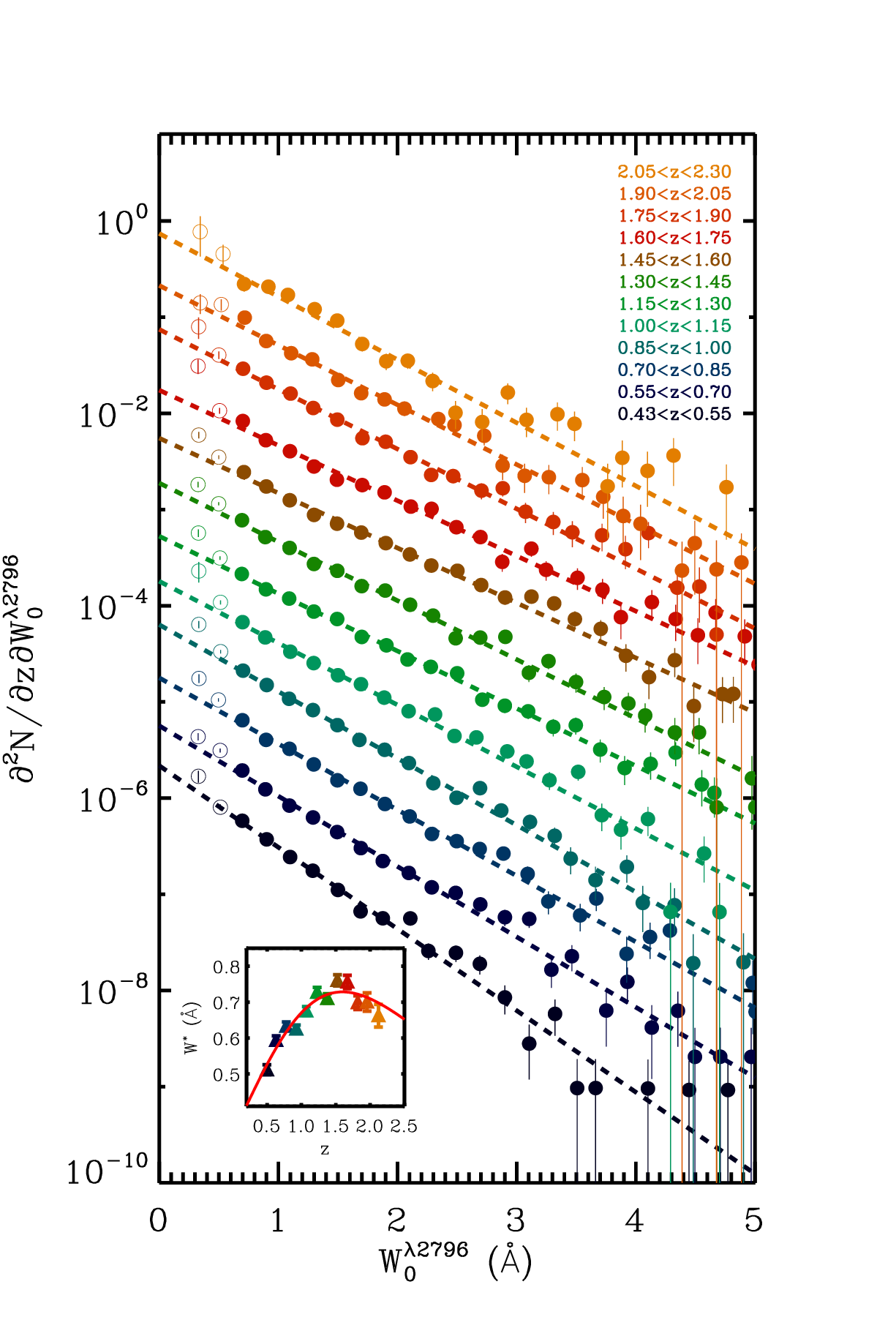}
\caption{Differential incidence rate \dndzdw\ of \mgii\ absorbers. For clarity, we have shifted the measurements by $-0.5~\mathrm{dex}$ from high to low redshift. The dashed lines represent best-fit exponential functions (Eq.~\ref{eq:d2ndwdz}) in the range $0.6$ \AA\ $<\rewmgiione<5$ \AA, i.e. including all the filled circles. Weaker absorbers follow a different distribution. The inset shows the redshift evolution of the $W^*$ parameter and the red solid line shows the best-fit parametrization (Eq.~\ref{eq:fit_sfr}).
}
\label{fig:dndzdw}
\end{figure}

We measure the incidence rate of the \mgii\ absorbers detected above. We estimate it using bins with $\Delta \rewmgiione=0.2$ \AA~ and $\Delta z = 0.15$. We start the lowest redshift bin at $z=0.43$ to avoid the region contaminated by Galactic \caii\ absorption and extend the highest redshift bin to $z=2.30$ to include highest-redshift absorbers. We present the measurements and Poisson errors in Figure \ref{fig:dndzdw}. For clarity, we have shifted  the measurements from high redshift to low redshift by $-0.5$ dex. The filled circles represent strong absorbers with $\rewmgiione>0.6$ \AA, while open circles indicate weak absorbers with $0.2$ \AA~$<\rewmgiione<0.6$ \AA. The rest equivalent width distributions are found to follow an exponential distribution at all redshifts. To summarize the overall dependence we perform a least-square fit  to all \dndzdw~data points with $0.6$ \AA~$<\rewmgiione<5.0$ \AA\ using an exponential function form:
\begin{equation}
\frac{\partial^2 N}{\partial z \partial \rewmgiione} (z,\rewmgiione) = \frac{N^*(z)}{W^*(z)}e^{-\frac{\rewmgiione}{W^*(z)}} \mathrm{.}
\label{eq:d2ndwdz}
\end{equation}
We present the best-fit parameters $N^*$ and $W^*$ in Table \ref{tbl:nw} and show $W^*$ in the inset in Figure \ref{fig:dndzdw}.  We also show the best-fit relations as dashed lines in the figure.  
It is remarkable to see that the simple dependence given in Eq.~\ref{eq:d2ndwdz} is able to describe 240 independent data points. Our fitting process does not include weak absorbers (shown as open circles) which appear to be drawn from a different distribution. The extrapolation of the exponentials clearly underestimates the incidence rate of such systems. The scale factor of the exponential form, $W^*$, is found to have a strong redshift dependence. It increases up to $z\sim1.5$ and decreases beyond. We now investigate this redshift evolution in more detail.

\input{tablenw.tex}

\begin{figure*}
\epsscale{1.2}
\plotone{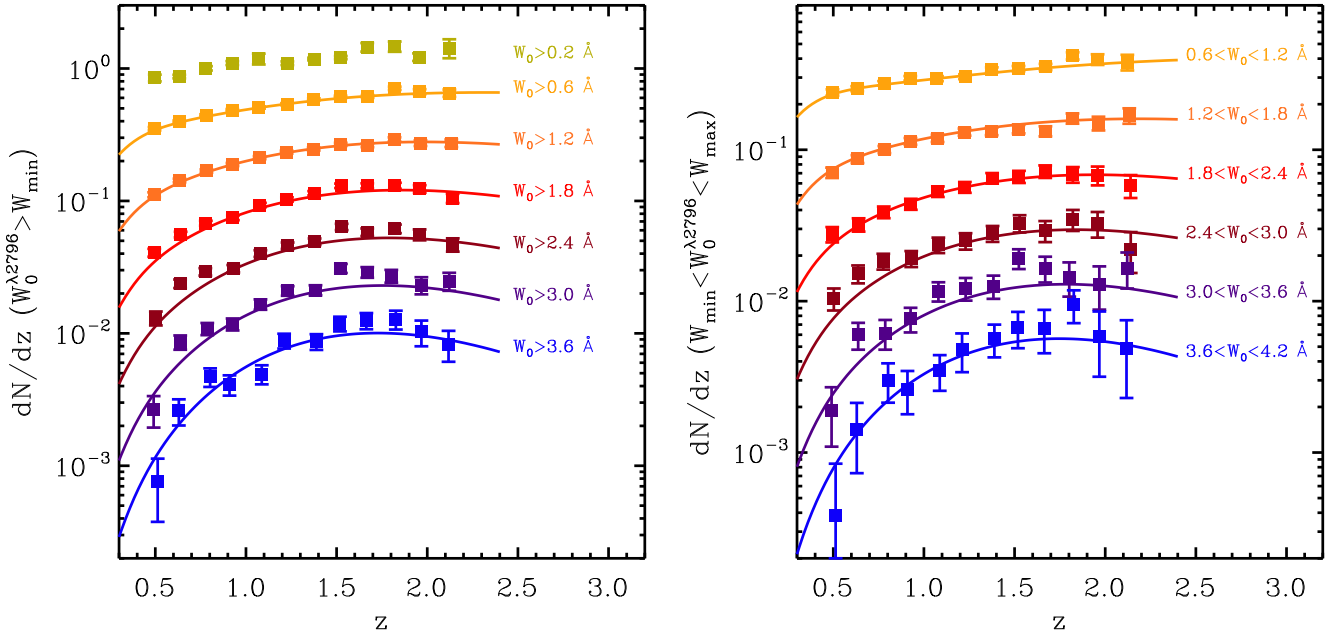}
\caption{
Cumulative (left) and differential (right) incidence rates \dndz\ of \mgii\ absorbers. The solid lines show the best-fit parametrization (Eq.~\ref{eq:fit_sfr}). The redshift evolution is much stronger for stronger absorbers.
}
\label{fig:cumdndzdw}
\end{figure*}

\subsection{The redshift evolution of \mgii\ absorbers}\label{sec:dndzdwevolv}

\begin{figure*}
\epsscale{0.8}
\plotone{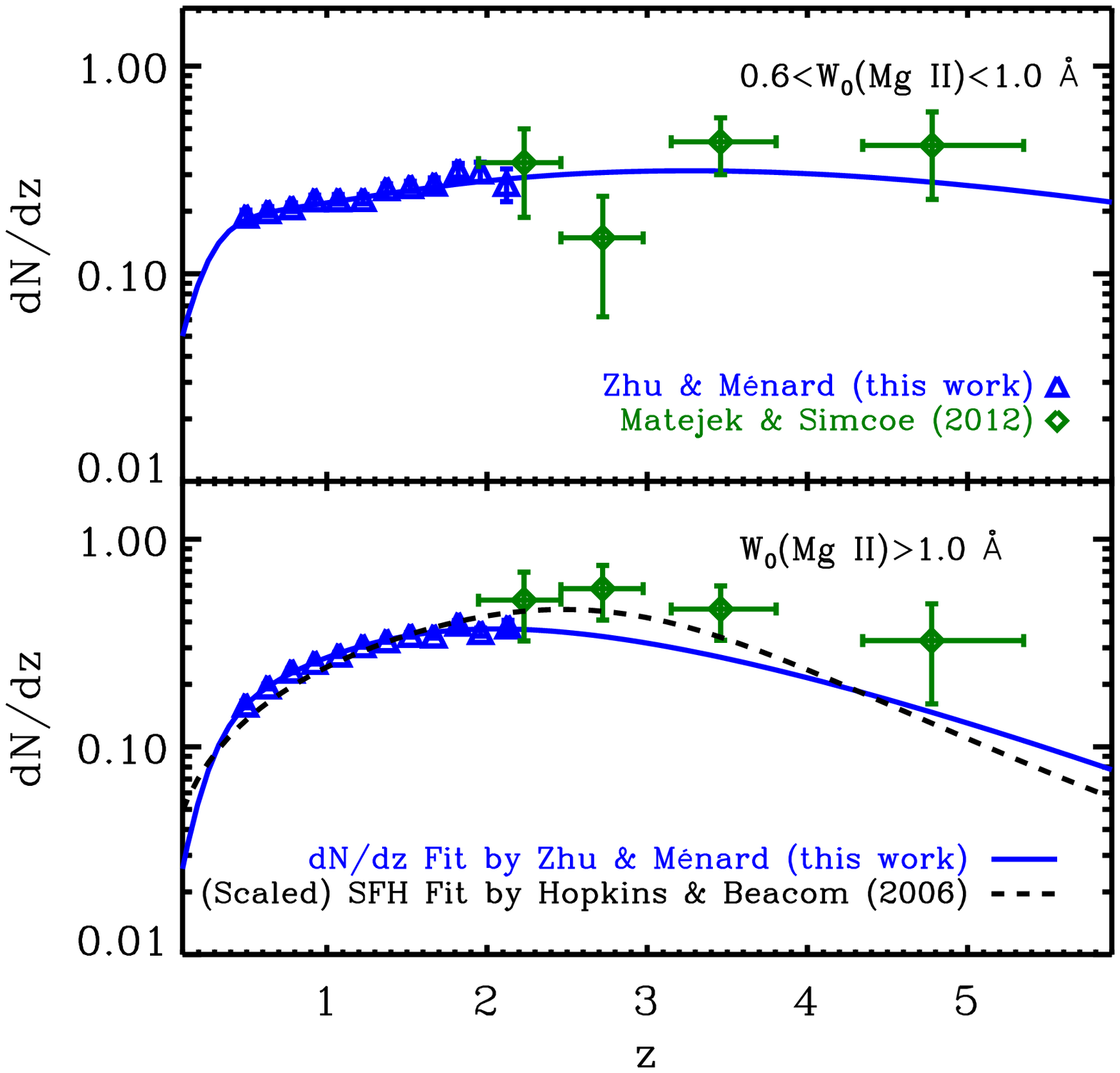}
\caption{
Incidence rate \dndz\ of \mgii\ absorbers with $0.6$ \AA\ $<\rewmgiione<1.0$ \AA\ (\emph{top}) and $\rewmgiione>1.0$ \AA~(\emph{bottom}). The blue points 
are from the present study based on SDSS data
are our measurements at moderate redshifts, 
while the green points show near-infrared measurements from \citet{matejek12a}. The blue solid lines show the global fit of Eq.~\ref{eq:fit_sfr} to all \dndzdw\ data points.
 In the lower panel, the dashed line shows the best-fit cosmic star formation history by \citet{hopkinsAM06a}, scaled to match $dN/dz$ \AA\ at $z\sim1.5$.
}
\label{fig:cumdndzdwallz}
\end{figure*}

We define the incidence rate of \mgii\ absorbers in a given range of rest equivalent width as
\begin{eqnarray}
\frac{dN}{dz} (W_{\mathrm{min}}<W_0<W_{\mathrm{max}}) &=&
\int^{W_\mathrm{max}}_{W_\mathrm{min}} \frac{\partial^2 N}{\partial z \partial W_0} d W_0~.
\end{eqnarray}
In Figure \ref{fig:cumdndzdw}, we present this quantity as well as the cumulative incidence rates above a given rest equivalent width, as a function of redshift. The incidence of absorbers with $\rewmgiione\sim0.6$ \AA\ increases  by less than a factor $2$ between $z=0.5$ and $z=2$. In contrast, stronger absorbers experience a stronger redshift evolution: from $z=0.5$ to $z=1.5$ the incidence rate of absorbers with $\rewmgiione>3$ \AA~increases by about an order of  magnitude. Interestingly, their incidence rate then flattens out from $z=1.5$  to $z\sim2$ and decreases towards higher redshift.

To characterize the redshift evolution of \mgii\ absorbers over a broader range of redshift we include recent incidence rate measurements by \citet{matejek12a}. Using near-infrared data these authors have estimated the incidence rate of \mgii\ absorbers up to $z=5.5$. The combined results are presented in Figure~\ref{fig:cumdndzdwallz}. Over the entire redshift range $0.4\lesssim z\lesssim5.5$, i.e., about 60\% of the age of the Universe, the incidence rate of weaker absorbers (with $0.6$ \AA~$<\rewmgiione<1.0$ \AA) is roughly the same, within a factor of 2. In contrast, the incidence rate for stronger absorbers (with $\rewmgiione>1.0$ \AA) increases up to $z\sim2$ and then decreases up to $z\sim5$. For those systems, the global redshift evolution of the incidence rate appears to be very similar to the cosmic star formation history \citep[SFH, \eg][]{hopkinsAM06a, zhu09a}. This is illustrated by the dashed line in the lower panel of Figure \ref{fig:cumdndzdwallz} which shows the best-fit cosmic SFH by \citet{hopkinsAM06a}. The amplitude of this curve has been scaled to match the amplitude of the \mgii\ incidence rate at $z\sim1.5$. The overall shapes of these two quantities are strikingly similar, pointing to a direct connection between strong absorbers and star formation. To quantify this further, we introduce a new parametrization of the incidence rate of \mgii\ absorbers combining constraints from our SDSS-based results and higher-redshift measurements from \citet{matejek12a}. We choose a functional form inspired by the commonly-used one for the cosmic SFH \citep[\eg][]{cole01a, hopkinsAM06a}:
\begin{equation}
\frac{\partial^2 N}{\partial z \partial \rewmgiione} (\rewmgiione, z) = g(z)~e^{-\frac{\rewmgiione}{W^*(z)}} \mathrm{,}
\label{eq:fit_sfr}
\end{equation}
\noindent where
\begin{equation}
g(z) = g_0~\frac{(1+z)^{\alpha_g}}{1+(\frac{z}{z_g})^{\beta_g}} \mathrm{,}\nonumber
\end{equation}
\noindent and
\begin{equation}
W^*(z) = W^*_0~\frac{(1+z)^{\alpha_W}}{1+(\frac{z}{z_W})^{\beta_W}} \mathrm{,}\nonumber
\end{equation}
\noindent in which $\alpha_{g}$, $\alpha_{W}$, $\beta_{g}$, $\beta_{W}$, $z_{g}$, and $z_{W}>0$. 
We perform a global least-squares fit to all \dndzdw~measurements at $0.6$ \AA~$<\rewmgiione<5.0\,{\rm\AA}$ at all redshifts with the parametrization above.  
We note that given their large error bars, the near-infrared high-redshift measurements contribute only weakly to the fit.
The constraints are dominated by the more precise measurements presented in this work.  
The best-fit parameters and their formal errors are given in Table \ref{tbl:z} and the 
corresponding incidence rates are shown with solid lines in Figures~\ref{fig:cumdndzdw}  and \ref{fig:cumdndzdwallz}. In both cases the parametrization given in Eq.~\ref{eq:fit_sfr} is an accurate representation of the data points over the entire redshift range. In addition we accurately reproduce the redshift evolution of $W^*(z)$, introduced in Eq.~\ref{eq:d2ndwdz}, as shown in the inset of  Figure \ref{fig:dndzdw}.

\input{tablez.tex}

\subsection{Discussion}

Using about 35,000 intervening \mgii\ absorbers from the SDSS and near-infrared data from \citet{matejek12a} we have shown that the evolution of the incidence rate of strong \mgii\ absorbers is very similar to that of the cosmic SFH over the entire range $0.4<z<5.5$. This is in agreement with previous results \citep[\eg][]{bergeron91a, nestor05a, prochter06a} but now shown with a much higher precision. 

Several studies have suggested a connection between strong \mgii\ absorbers and star formation. \citet{bergeron91a} showed that most galaxies identified with \mgii\ absorbers in their sample are fairly blue and show \oii~emission.  \citet{norman96a} detected strong \mgii\ absorption arising from gas around starburst galaxy NGC $520$.  Using near-infrared integral field spectroscopy, \citet{bouche07a} detected strong \ha~emission around $14$ out of $21$ strong \mgii\ absorbers with $\rewmgiione>2$ \AA.  \citet{nestor11a} studied galaxies around two strong \mgii\ absorbers with $\rewmgiione>3$ \AA, and found that they are both associated with bright emission-line galaxies with large specific star formation rate for their masses. \citet{menard11a} showed that the mean \oii\ luminosity density traced by a sample of about $8,500$ \mgii\ absorbers from the SDSS follows that of the cosmic SFH.
A number of studies of galaxy spectra also support the connection between absorbers and star formation.  \citet{tremonti07a} detected \mgii\ outflows in $10$ out of $14$ post-starburst galaxies. Using galaxy spectra from the DEEP2 survey, \citet{weiner09a} showed that blue-shifted \mgii\ absorption is ubiquitous in star-forming galaxies at $z\sim1.4$, and the \mgii\ equivalent width and outflow velocity increase with stellar mass and star formation rate.  \citet{rubin10a} extended the analysis to lower redshift at $0.7<z<1.5$ and reached a similar conclusion.
More recently, using stacked spectra of background galaxies, \citet{bordoloi11a} studied the radial and azimuthal distribution of \mgii\ gas of galaxies at $0.5<z<0.9$.  They showed that blue galaxies have a significantly larger average \mgii\ equivalent width at close galactocentric radii than red galaxies. They also showed that the average \mgii\ equivalent width is larger at larger azimuthal angle, indicating the presence of a strongly bipolar outflow aligned with the disk rotation axis.
Our results further support the connection between strong \mgii\ absorbers and star formation, across a wide range of redshifts.

\section{Summary}\label{sec:summary}

The \mgiidoublet\ absorption line doublet probes low-ionization and neutral gas in the Universe.  
We have developed a generic and fully-automatic algorithm to detect absorption lines in the spectra of astronomical sources. The estimation of the flux continuum is based on nonnegative matrix factorization (NMF), a vector decomposition technique similar to principal component analysis (PCA) but with the additional requirement of nonnegativity.
We then applied this algorithm to a sample of about 100,000 quasar spectra from the SDSS DR7 dataset. Our results are summarized as follows:
\begin{itemize}
\item We detected $40,429$ \mgii\ absorbers, with $35,752$ intervening systems, defined as
$z_{\rm abs}<z_{\rm QSO}-0.04$, corresponding to a $\Delta v>\sim12,000~\kms$. This doubles the size of previously published \mgii\ catalogs. The dataset is available at:\\ 
 {\tt {\url{http://www.pha.jhu.edu/~gz323/jhusdss}}}.\\ 
 Future updates including new data releases can be found at the same address.
\item We determined the completeness and purity of our line detection algorithm and validated it with the visually-inspected Pittsburgh \mgii\ catalog \citep[][based on the SDSS DR4 subset]{quider11a}.
\item We measured the differential incidence rate \dndzdw\ of \mgii\ absorbers: the rest equivalent width distribution of systems with $W_0>0.6\,{\rm \AA}$ is well-represented  by an exponential at all redshifts. The shape of this distribution changes for weaker absorbers.
Combining our SDSS-based results and near-infrared measurements of \mgii\ absorbers by 
\citet{matejek12a} we introduced a new parametrization of  the differential incidence rate \dndzdw\ of \mgii\ absorbers (Eq.~\ref{eq:fit_sfr}), valid over the entire redshift range $0.4<z<5.5$.
\item Over this entire redshift range, which covers more than 60\% of the age of the Universe, the incidence rate evolution of strong absorbers (with $W_0>1\,{\rm \AA})$ is strikingly similar to the cosmic star formation history, suggesting a direct link between these two quantities.
\end{itemize}

The algorithm presented in this work is generic and can easily be used in other contexts. It is not limited to quasars but can estimate the continuum flux of any ensemble of sources, for example galaxies. It can also be used to detect any other line, in absorption or in emission and at any rest-frame wavelength. It is readily applicable to upcoming surveys such as eBOSS \citep[][]{comparat12a}, BigBOSS \citep[][]{schlegel09a}, and PFS \citep[][]{ellis12a}.

\acknowledgments

We thank Alex Szalay, Tamas Budavari, and Ani Thakar for sharing their computational resources. We have made extensive use of SDSS IDL libraries written by David  Schlegel, Michael Blanton, David Hogg, and others. We also acknowledge the usage of the MPFIT package written by Craig Markwardt. The authors acknowledge funding support from NSF grant AST-1109665 and the Alfred P. Sloan foundation. Funding for the SDSS and SDSS-II has been provided by the Alfred P. Sloan Foundation, the Participating Institutions, the National Science Foundation, the U.S. Department of Energy, the National Aeronautics and Space Administration, the Japanese Monbukagakusho, the Max Planck Society, and the Higher Education Funding Council for England. The SDSS Web Site is http://www.sdss.org/.

\bibliographystyle{apj}
\bibliography{zhuref.bib}

\begin{appendix}

\section{Construction of the NMF basis sets}

In order to create a basis set of eigenspectra we choose to use flux-normalized quasar spectra.
To do so we use four different wavelength ranges  where quasar spectra are relatively featureless. This choice is based on the median quasar spectral energy distribution (SED) given in \cite{vandenberk01a} and summarized in Table~\ref{tbl:norm}. For each normalization wavelength range, we normalize the observed spectra with the mean flux within the range. We create a basis set for each range using all quasars with the range fully covered in the spectra.   When fitting the continuum of a given quasar, we choose the basis set  of eigenspectra whose median redshift is closest to the quasar redshift.  This guarantees that each quasar is fit with a set of eigenspectra  that are built from the maximal number of available quasars covering the same wavelengths.  

\input{tablenorm.tex}

As an example, we show the NMF basis set of $12$ eigenspectra in the redshift bin with $0.4<z<1.8$, for the normalization wavelength range $3020-3100$ \AA. In the first five panels, we also label the prominent features such as permitted metal emission lines, forbidden lines, and Balmer series. The natural separation of different types of emission lines illustrates the power of the NMF vector decomposition to characterize quasar spectra.

\begin{figure*}
\epsscale{1.2}
\plotone{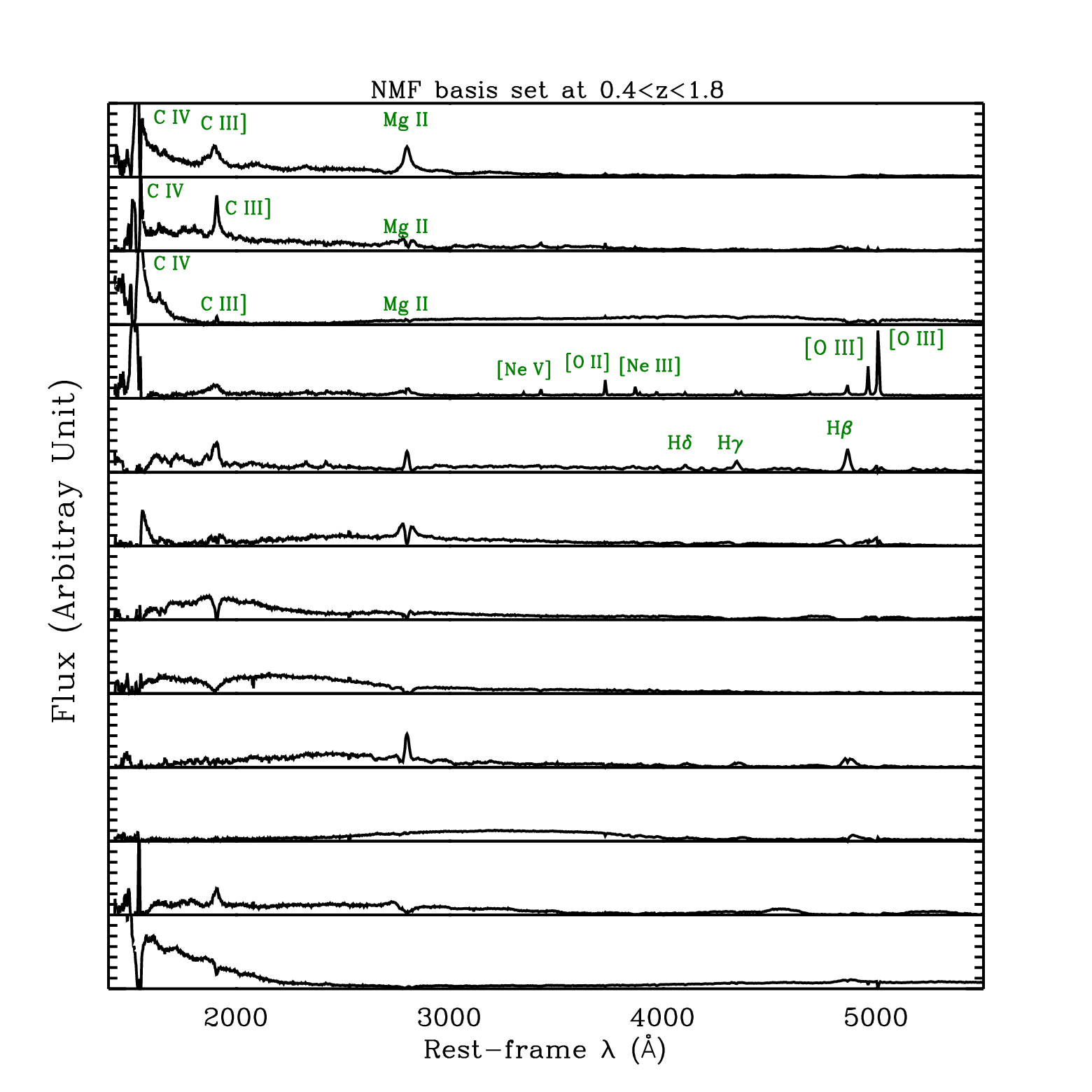}
\caption{The NMF basis set of eigenspectra at $0.4<z<1.8$. 
We label the permitted metal emission lines, forbidden lines,
and Balmer series in the first five panels.}
\label{fig:nmfbasis}
\end{figure*}

\section{Comparison with the Pittsburgh catalog}

\begin{figure*}
\epsscale{1.0}
\plotone{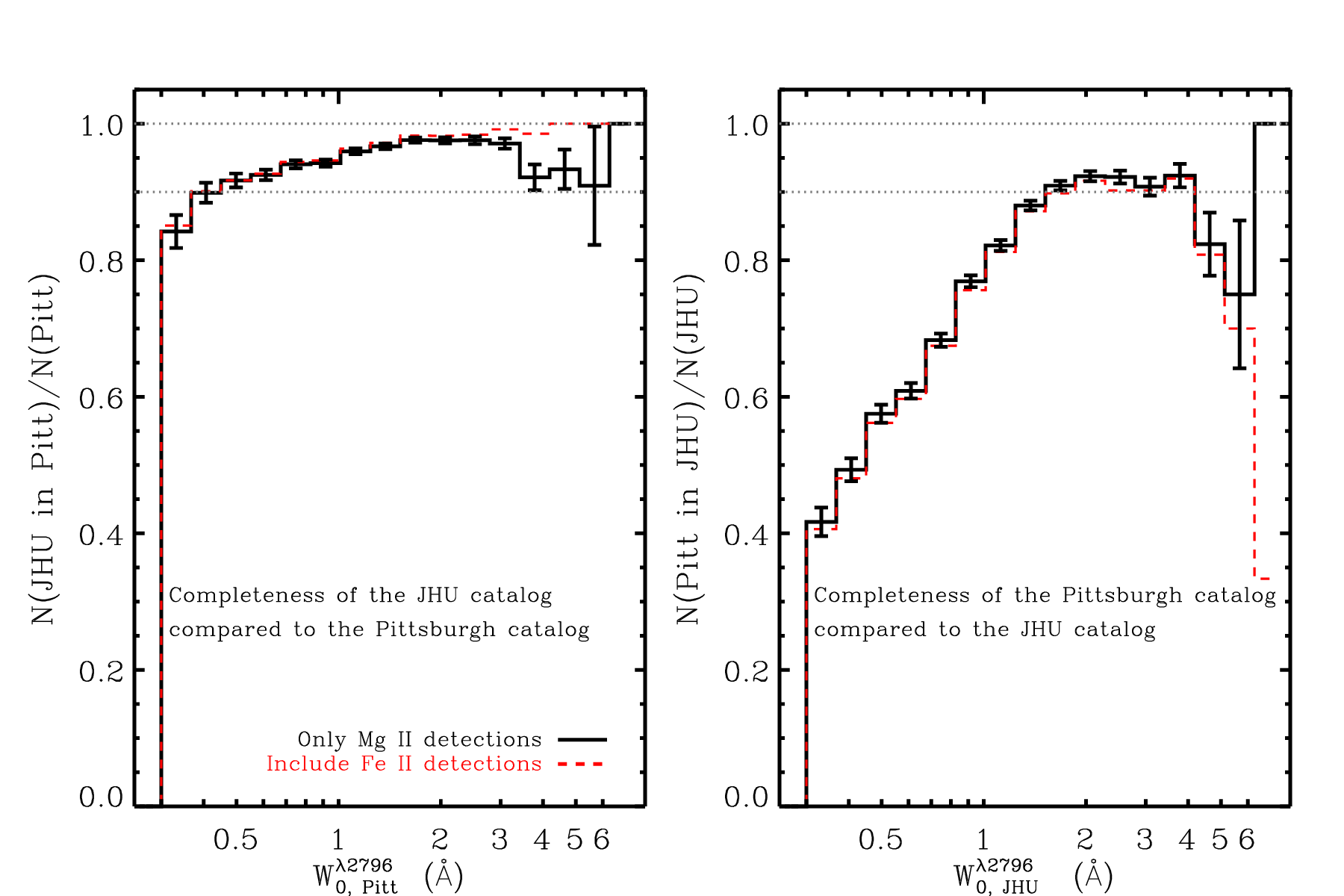}
\caption{
Comparison of the JHU and Pittsburgh catalogs. The black line in the left panel shows the fraction of \mgii\ absorbers reported by the Pittsburgh pipeline that are also detected by our pipeline.  To guide the eye, we overplot two horizontal dotted lines at $100\%$ and $90\%$.  Within the search window, we recovered  $14,079$ ($\sim95\%$) of $14,715$ absorbers reported in the Pittsburgh catalog. The missing absorbers are due to low $SNR$ or masked pixels and did not pass \criterionmgii. If we also include \feii\ absorbers that satisfy \criterionfeii,  we recovered close to $100\%$ of strong absorbers. In the right panel, we show the fraction of \mgii\ absorbers in the JHU catalog that are also included in the Pittsburgh catalog.  We detected $18,748$ in total, with $4669$ ($\sim25\%$) systems not  reported in the Pittsburgh catalog.  In Figure \ref{fig:jhunomatchstrong} and \ref{fig:jhunomatchcomposite},  we show that these extra absorbers we found are bonafide absorbers.}
\label{fig:ffcomppitts}
\end{figure*}

%
The Pittsburgh catalog \citep{quider11a,nestor05a} which uses visual inspection provides  a valuable reference to validate our pipeline.  Here we present  a detailed comparison between the two \mgii\ absorber catalogs.
There are $41,881$ common quasars searched for \mgii\ doublets in both surveys. Within the search window, we detected $18,748$ \mgii\ absorbers with  $\rewmgiione>0.02$ \AA, while the Pittsburgh pipeline led to $14,715$ objects.  In Figure \ref{fig:ffcomppitts}, we show the fraction of absorbers  detected in both surveys as a function of \rewmgiione.  In the left panel, we show that among the $14,715$ detected  by the Pittsburgh pipeline, we recovered $14,079$ ($\sim95\%$).  For the remaining $\sim5\%$ the noise level of the residuals at the location of the absorbers is too high for our \criterionmgii\ (Eq.~\ref{eq:criterionmgii}) to be satisfied. Such regions of the spectra are not included in our redshift path and these non-detections do not affect our statistical analysis.
If we include absorber systems that do not pass  \criterionmgii\ but satisfy
\criterionfeii\ (Eq.~\ref{eq:criterionfeii}), we recover close to $100\%$ of the strong absorbers detected by the Pittsburgh pipeline, as shown by the red line in the left panel of Figure~\ref{fig:ffcomppitts}.

\begin{figure*}
\epsscale{1.0}
\plotone{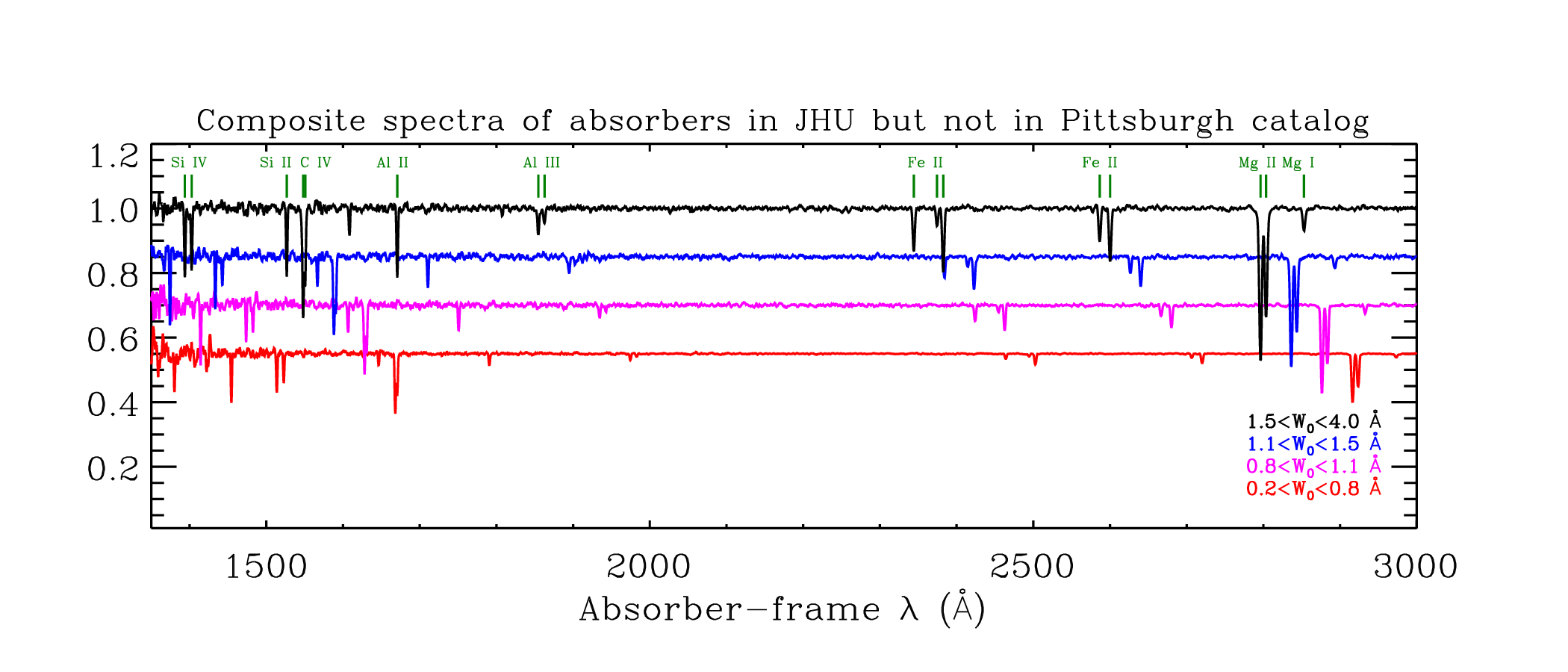}
\caption{
Composite spectra of absorbers detected by the JHU pipeline but not by the Pittsburgh pipeline. We divided the sample of absorbers with $\rewmgiione<4.0$ \AA\ into four subsamples with $0.2$ \AA~$<\rewmgiione<0.8$ \AA~
(the number of absorbers $N=2971$), 
$0.8$ \AA~$<\rewmgiione<1.1$ \AA~($N=831$),
$1.1$ \AA~$<\rewmgiione<1.5$ \AA~($N=469$),
and $1.5$ \AA~$<\rewmgiione<4.0$ \AA~($N=379$). 
We have labeled the locations of prominent absorption lines to guide the eye. The expected absorption lines show these systems are real absorbers.}
\label{fig:jhunomatchcomposite}
\end{figure*}

\begin{figure*}
\epsscale{1.0}
\plotone{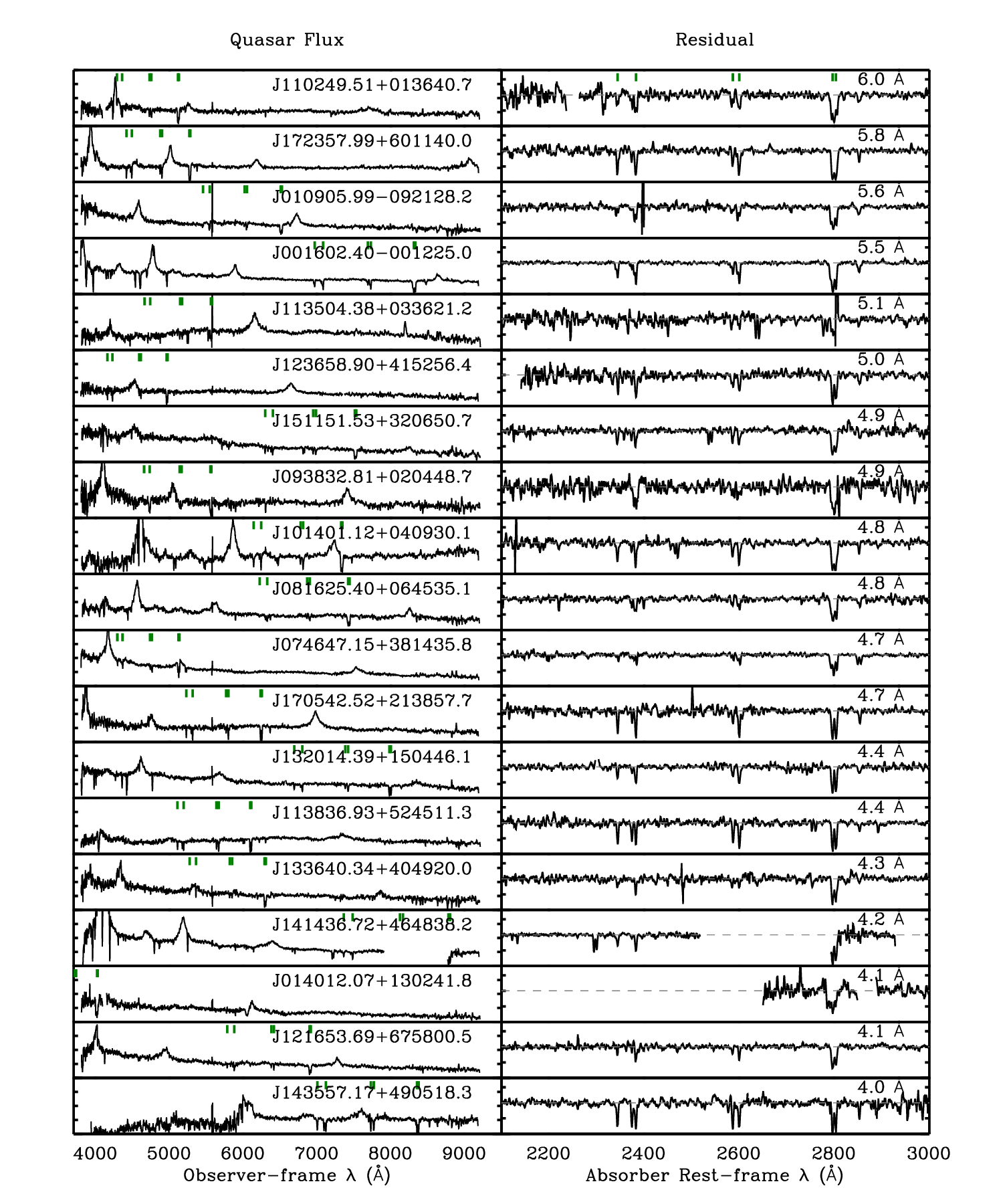}
\caption{Individual spectra of absorbers with $\rewmgiione>4.0$  \AA\ detected by the JHU pipeline but not by the Pittsburgh pipeline. In the left panels we show the observed quasar spectra in the observer frame, while in the right panels we show the final residuals in the absorber frame. We show the IAU-formatted names of the quasars in the left panels. We also label the locations of \mgiidoublet~and  \feii~$\lambda2600$, $\lambda2586$, $\lambda2383$, $\lambda2344$ lines. To guide the eye, we overplot a gray dashed line at unity in the  right panels. We show nearly all absorbers are real absorbers that can be confirmed with \feii\ lines.
}
\label{fig:jhunomatchstrong}
\end{figure*}

In the right panel of the figure, we show the fraction of absorbers of our catalog that are also detected by the Pittsburgh pipeline. It shows that $4669$ of the systems we detected are not reported in the Pittsburgh catalog. The fraction of missing systems is a function of rest equivalent width and increases for weaker systems. We now demonstrate that these absorbers are bonafide \mgii\ absorbers. For the strongest systems ($\rewmgiione>4\,{\rm \AA}$) we visually inspected the spectra and, based on the presence of additional metal lines, we were able to confirm the nature of the systems. This is illustrated in Figure~\ref{fig:jhunomatchstrong}, where we have labeled six strongest lines: \mgiidoublet, \feii~$\lambda2600$, $\lambda2586$, $\lambda2383$, and $\lambda2344$. Nearly all \mgii\ absorbers can be confirmed with the \feii~lines at the right locations. For weaker absorbers, we construct composite spectra. We divide the sample of absorbers missing in the Pittsburgh catalog into four subsamples with $0.2$ \AA~$<\rewmgiione<0.8$ \AA, $0.8$ \AA~$<\rewmgiione<1.1$ \AA, $1.1$ \AA~$<\rewmgiione<1.5$ \AA, and $1.5$ \AA~$<\rewmgiione<4.0$ \AA. The numbers of absorbers in each subsample are  $2971$, $831$, $469$, and $379$, respectively. The composite spectra shown in Figure~\ref{fig:jhunomatchcomposite} display all the expected metal absorption lines, which shows that the absorbers detected by our pipeline but not reported in the Pittsburgh catalog are real \mgii\ absorption-line systems. This confirms that our pipeline leads to robust detections of \mgii\ absorbers.

\end{appendix}
 
\end{document}

%% file: definition.tex
\def\eg{e.g., }

\def\oii{[\ion{O}{2}]}

\def\mgii{\ion{Mg}{2}}
\def\mgiidoublet{\ion{Mg}{2}~$\lambda\lambda2796,2803$}

\def\feii{\ion{Fe}{2}}

\def\ciii{\ion{C}{3}}
\def\civ{\ion{C}{4}}

\def\oi{\ion{O}{1}}

\def\caii{\ion{Ca}{2}}
\def\criterionmgii{{\tt Criterion\_MgII}}
\def\criterionfeii{{\tt Criterion\_FeII}}
\def\caiidoublet{\ion{Ca}{2}~$\lambda\lambda3934,3969$}

\def\ha{H$\alpha$}

\newcommand{\kms}{\ensuremath{{\rm km~s}^{-1}}}

\newcommand{\rewmgiione}{\ensuremath{W_0^{\lambda2796}}}
\newcommand{\rewmgiitwo}{\ensuremath{W_0^{\lambda2803}}}
\newcommand{\dndzdw}{\ensuremath{\partial^2N/\partial z\partial \rewmgiione}}
\newcommand{\dndz}{\ensuremath{dN/dz}}

%% file: tablenw.tex
\begin{deluxetable}{cccc}
\tabletypesize{\scriptsize}
\tablecolumns{4}
\tablecaption{Best-fit parameters of Equation~\ref{eq:d2ndwdz}}
\tablehead{
 \colhead{$z$} &  \colhead{$<z>$} &  \colhead{$N^*$} &  \colhead{$W^*$} 
}
\startdata
\input{tablenwdata.tex}
\enddata
\label{tbl:nw}
\end{deluxetable}

%% file: tablenwdata.tex
$0.43-0.55$ & $0.48$ & $1.11\pm0.04$ & $0.51\pm0.01$ \\
$0.55-0.70$ & $0.63$ & $1.06\pm0.03$ & $0.59\pm0.01$ \\
$0.70-0.85$ & $0.78$ & $1.13\pm0.03$ & $0.63\pm0.01$ \\
$0.85-1.00$ & $0.93$ & $1.25\pm0.03$ & $0.63\pm0.01$ \\
$1.00-1.15$ & $1.08$ & $1.21\pm0.03$ & $0.68\pm0.01$ \\
$1.15-1.30$ & $1.23$ & $1.22\pm0.03$ & $0.73\pm0.01$ \\
$1.30-1.45$ & $1.38$ & $1.34\pm0.03$ & $0.71\pm0.01$ \\
$1.45-1.60$ & $1.53$ & $1.33\pm0.04$ & $0.76\pm0.02$ \\
$1.60-1.75$ & $1.68$ & $1.32\pm0.05$ & $0.76\pm0.02$ \\
$1.75-1.90$ & $1.83$ & $1.65\pm0.06$ & $0.70\pm0.02$ \\
$1.90-2.05$ & $1.98$ & $1.49\pm0.08$ & $0.70\pm0.03$ \\
$2.05-2.30$ & $2.13$ & $1.55\pm0.10$ & $0.66\pm0.03$ \\

%% file: tablez.tex
\begin{deluxetable}{cccc}
\tabletypesize{\scriptsize}
\tablecolumns{4}
\tablecaption{Best-fit parameters of equation $5$}
\tablehead{
\colhead{$g_0$} & \colhead{$\alpha_g$} & \colhead{$z_g$} & \colhead{$\beta_g$}
}
\startdata
\input{tablezdata.tex}
\enddata
\label{tbl:z}
\end{deluxetable}

%% file: tablezdata.tex
\vspace{0.06cm}
$0.63\pm0.39$ & $5.38\pm1.08$ & $0.41\pm0.06$ & $2.97\pm0.59$ \\ \hline
\hline
$W_0$ & $\alpha_W$ & $z_W$ & $\beta_W$ \\ \hline
$0.33\pm0.03$ & $1.21\pm0.19$ & $2.24\pm0.28$ & $2.43\pm0.25$ \\

%% file: tablenorm.tex
\begin{deluxetable*}{ccc}
\tabletypesize{\scriptsize}
\tablecolumns{3}
\tablecaption{Normalization schemes used in the NMF fitting}
\tablehead{
\colhead{Normalization wavelength range\tablenotemark{a}} & 
\colhead{Eigenspectra construction redshift range\tablenotemark{b}} & 
\colhead{Continuum fitting redshift range\tablenotemark{c}}
}
\startdata   
$4150~{\rm \AA}-4250~{\rm \AA}$ & $z<1.0$      & $z<0.6$       \\
$3020~{\rm \AA}-3100~{\rm \AA}$ & $0.4<z<1.8$  & $0.6<z<1.0$   \\
$2150~{\rm \AA}-2250~{\rm \AA}$ & $0.8<z<2.8$  & $1.0<z<2.5$   \\
$1420~{\rm \AA}-1500~{\rm \AA}$ & $2.0<z<4.8$  & $2.5<z<4.7$   \\
\enddata
\tablenotetext{a}{When constructing the basis set of eigenspectra, 
we choose to work on flux-normalized quasar spectra.
We choose these normalization wavelength ranges where quasar spectra are relatively featureless.}
\tablenotetext{b}{The redshift ranges where the normalization wavelength 
range is covered by SDSS.  we make use of all available 
quasars within each range to construct the basis set of eigenspectra.}
\tablenotetext{c}{For quasars within these redshift ranges, we fit their 
observed spectra with the basis set of eigenspectra constructed with quasars in the second column. 
These ranges are chosen so that each quasar is fit with the basis 
set of eigenspectra constructed using maximal number of quasars that 
cover the same wavelengths.}
\label{tbl:norm}
\end{deluxetable*}

%% file: ms.bbl
\begin{thebibliography}{52}
\expandafter\ifx\csname natexlab\endcsname\relax\def\natexlab#1{#1}\fi

\bibitem[{{Abazajian} {et~al.}(2009){Abazajian}, {Adelman-McCarthy},
  {Ag{\"u}eros}, {Allam}, {Allende Prieto}, {An}, {Anderson}, {Anderson},
  {Annis}, {Bahcall}, \& et~al.}]{abazajian09a}
{Abazajian}, K.~N., {Adelman-McCarthy}, J.~K., {Ag{\"u}eros}, M.~A., {et~al.}
  2009, \apjs, 182, 543

\bibitem[{{Bergeron} \& {Boiss{\'e}}(1991)}]{bergeron91a}
{Bergeron}, J., \& {Boiss{\'e}}, P. 1991, \aap, 243, 344

\bibitem[{{Blanton} \& {Roweis}(2007)}]{blanton07a}
{Blanton}, M.~R., \& {Roweis}, S. 2007, \aj, 133, 734

\bibitem[{{Bordoloi} {et~al.}(2011){Bordoloi}, {Lilly}, {Knobel}, {Bolzonella},
  {Kampczyk}, {Carollo}, {Iovino}, {Zucca}, {Contini}, {Kneib}, {Le Fevre},
  {Mainieri}, {Renzini}, {Scodeggio}, {Zamorani}, {Balestra}, {Bardelli},
  {Bongiorno}, {Caputi}, {Cucciati}, {de la Torre}, {de Ravel}, {Garilli},
  {Kova{\v c}}, {Lamareille}, {Le Borgne}, {Le Brun}, {Maier}, {Mignoli},
  {Pello}, {Peng}, {Perez Montero}, {Presotto}, {Scarlata}, {Silverman},
  {Tanaka}, {Tasca}, {Tresse}, {Vergani}, {Barnes}, {Cappi}, {Cimatti},
  {Coppa}, {Diener}, {Franzetti}, {Koekemoer}, {L{\'o}pez-Sanjuan},
  {McCracken}, {Moresco}, {Nair}, {Oesch}, {Pozzetti}, \&
  {Welikala}}]{bordoloi11a}
{Bordoloi}, R., {Lilly}, S.~J., {Knobel}, C., {et~al.} 2011, \apj, 743, 10

\bibitem[{{Bouch{\'e}} {et~al.}(2006){Bouch{\'e}}, {Murphy}, {P{\'e}roux},
  {Csabai}, \& {Wild}}]{bouche06a}
{Bouch{\'e}}, N., {Murphy}, M.~T., {P{\'e}roux}, C., {Csabai}, I., \& {Wild},
  V. 2006, \mnras, 371, 495

\bibitem[{{Bouch{\'e}} {et~al.}(2007){Bouch{\'e}}, {Murphy}, {P{\'e}roux},
  {Davies}, {Eisenhauer}, {F{\"o}rster Schreiber}, \& {Tacconi}}]{bouche07a}
{Bouch{\'e}}, N., {Murphy}, M.~T., {P{\'e}roux}, C., {et~al.} 2007, \apjl, 669,
  L5

\bibitem[{{Caulet}(1989)}]{caulet89a}
{Caulet}, A. 1989, \apj, 340, 90

\bibitem[{{Chelouche} \& {Bowen}(2010)}]{chelouche10a}
{Chelouche}, D., \& {Bowen}, D.~V. 2010, \apj, 722, 1821

\bibitem[{{Chen} {et~al.}(2010{\natexlab{a}}){Chen}, {Helsby}, {Gauthier},
  {Shectman}, {Thompson}, \& {Tinker}}]{chen10a}
{Chen}, H.-W., {Helsby}, J.~E., {Gauthier}, J.-R., {et~al.} 2010{\natexlab{a}},
  \apj, 714, 1521

\bibitem[{{Chen} {et~al.}(2010{\natexlab{b}}){Chen}, {Wild}, {Tinker},
  {Gauthier}, {Helsby}, {Shectman}, \& {Thompson}}]{chen10b}
{Chen}, H.-W., {Wild}, V., {Tinker}, J.~L., {et~al.} 2010{\natexlab{b}}, \apjl,
  724, L176

\bibitem[{{Churchill} {et~al.}(2000{\natexlab{a}}){Churchill}, {Mellon},
  {Charlton}, {Jannuzi}, {Kirhakos}, {Steidel}, \& {Schneider}}]{churchill00a}
{Churchill}, C.~W., {Mellon}, R.~R., {Charlton}, J.~C., {et~al.}
  2000{\natexlab{a}}, \apjs, 130, 91

\bibitem[{{Churchill} {et~al.}(2000{\natexlab{b}}){Churchill}, {Mellon},
  {Charlton}, {Jannuzi}, {Kirhakos}, {Steidel}, \& {Schneider}}]{churchill00b}
---. 2000{\natexlab{b}}, \apj, 543, 577

\bibitem[{{Churchill} {et~al.}(1999){Churchill}, {Rigby}, {Charlton}, \&
  {Vogt}}]{churchill99a}
{Churchill}, C.~W., {Rigby}, J.~R., {Charlton}, J.~C., \& {Vogt}, S.~S. 1999,
  \apjs, 120, 51

\bibitem[{{Cole} {et~al.}(2001){Cole}, {Norberg}, {Baugh}, {Frenk},
  {Bland-Hawthorn}, {Bridges}, {Cannon}, {Colless}, {Collins}, {Couch},
  {Cross}, {Dalton}, {De Propris}, {Driver}, {Efstathiou}, {Ellis},
  {Glazebrook}, {Jackson}, {Lahav}, {Lewis}, {Lumsden}, {Maddox}, {Madgwick},
  {Peacock}, {Peterson}, {Sutherland}, \& {Taylor}}]{cole01a}
{Cole}, S., {Norberg}, P., {Baugh}, C.~M., {et~al.} 2001, \mnras, 326, 255

\bibitem[{{Comparat} {et~al.}(2012){Comparat}, {Kneib}, {Escoffier}, {Zoubian},
  {Ealet}, {Lamareille}, {Mostek}, {Steele}, {Aubourg}, {Bailey}, {Bolton},
  {Brownstein}, {Dawson}, {Ge}, {Ilbert}, {Leauthaud}, {Maraston}, {Percival},
  {Ross}, {Schimd}, {Schlegel}, {Schneider}, {Thomas}, {Tinker}, \&
  {Weaver}}]{comparat12a}
{Comparat}, J., {Kneib}, J.-P., {Escoffier}, S., {et~al.} 2012, \mnras, 104

\bibitem[{{Connolly} {et~al.}(1995){Connolly}, {Szalay}, {Bershady}, {Kinney},
  \& {Calzetti}}]{connolly95a}
{Connolly}, A.~J., {Szalay}, A.~S., {Bershady}, M.~A., {Kinney}, A.~L., \&
  {Calzetti}, D. 1995, \aj, 110, 1071

\bibitem[{{Ellis} {et~al.}(2012){Ellis}, {Takada}, {Aihara}, {Arimoto},
  {Bundy}, {Chiba}, {Cohen}, {Dore}, {Greene}, {Gunn}, {Heckman}, {Hirata},
  {Ho}, {Kneib}, {Le Fevre}, {Murayama}, {Nagao}, {Ouchi}, {Seiffert},
  {Silverman}, {Sodre}, {Spergel}, {Strauss}, {Sugai}, {Suto}, {Takami},
  {Wyse}, \& {the PFS Team}}]{ellis12a}
{Ellis}, R., {Takada}, M., {Aihara}, H., {et~al.} 2012, ArXiv e-prints

\bibitem[{{Hewett} \& {Wild}(2010)}]{hewett10a}
{Hewett}, P.~C., \& {Wild}, V. 2010, \mnras, 405, 2302

\bibitem[{{Hopkins} \& {Beacom}(2006)}]{hopkinsAM06a}
{Hopkins}, A.~M., \& {Beacom}, J.~F. 2006, \apj, 651, 142

\bibitem[{{Kacprzak} \& {Churchill}(2011)}]{kacprzak11c}
{Kacprzak}, G.~G., \& {Churchill}, C.~W. 2011, \apjl, 743, L34

\bibitem[{{Kacprzak} {et~al.}(2011{\natexlab{a}}){Kacprzak}, {Churchill},
  {Barton}, \& {Cooke}}]{kacprzak11a}
{Kacprzak}, G.~G., {Churchill}, C.~W., {Barton}, E.~J., \& {Cooke}, J.
  2011{\natexlab{a}}, \apj, 733, 105

\bibitem[{{Kacprzak} {et~al.}(2010){Kacprzak}, {Churchill}, {Ceverino},
  {Steidel}, {Klypin}, \& {Murphy}}]{kacprzak10a}
{Kacprzak}, G.~G., {Churchill}, C.~W., {Ceverino}, D., {et~al.} 2010, \apj,
  711, 533

\bibitem[{{Kacprzak} {et~al.}(2011{\natexlab{b}}){Kacprzak}, {Churchill},
  {Evans}, {Murphy}, \& {Steidel}}]{kacprzak11b}
{Kacprzak}, G.~G., {Churchill}, C.~W., {Evans}, J.~L., {Murphy}, M.~T., \&
  {Steidel}, C.~C. 2011{\natexlab{b}}, \mnras, 416, 3118

\bibitem[{{Lanzetta} {et~al.}(1987){Lanzetta}, {Turnshek}, \&
  {Wolfe}}]{lanzetta87a}
{Lanzetta}, K.~M., {Turnshek}, D.~A., \& {Wolfe}, A.~M. 1987, \apj, 322, 739

\bibitem[{{Lee} \& {Seung}(1999)}]{lee99a}
{Lee}, D.~D., \& {Seung}, H.~S. 1999, \nat, 401, 788

\bibitem[{{Lundgren} {et~al.}(2009){Lundgren}, {Brunner}, {York}, {Ross},
  {Quashnock}, {Myers}, {Schneider}, {Al Sayyad}, \& {Bahcall}}]{lundgren09a}
{Lundgren}, B.~F., {Brunner}, R.~J., {York}, D.~G., {et~al.} 2009, \apj, 698,
  819

\bibitem[{{Matejek} \& {Simcoe}(2012)}]{matejek12a}
{Matejek}, M.~S., \& {Simcoe}, R.~A. 2012, ArXiv e-prints

\bibitem[{{M{\'e}nard} {et~al.}(2011){M{\'e}nard}, {Wild}, {Nestor}, {Quider},
  {Zibetti}, {Rao}, \& {Turnshek}}]{menard11a}
{M{\'e}nard}, B., {Wild}, V., {Nestor}, D., {et~al.} 2011, \mnras, 417, 801

\bibitem[{{Nestor} {et~al.}(2011){Nestor}, {Johnson}, {Wild}, {M{\'e}nard},
  {Turnshek}, {Rao}, \& {Pettini}}]{nestor11a}
{Nestor}, D.~B., {Johnson}, B.~D., {Wild}, V., {et~al.} 2011, \mnras, 412, 1559

\bibitem[{{Nestor} {et~al.}(2005){Nestor}, {Turnshek}, \& {Rao}}]{nestor05a}
{Nestor}, D.~B., {Turnshek}, D.~A., \& {Rao}, S.~M. 2005, \apj, 628, 637

\bibitem[{{Norman} {et~al.}(1996){Norman}, {Bowen}, {Heckman}, {Blades}, \&
  {Danly}}]{norman96a}
{Norman}, C.~A., {Bowen}, D.~V., {Heckman}, T., {Blades}, C., \& {Danly}, L.
  1996, \apj, 472, 73

\bibitem[{{Prochter} {et~al.}(2006){Prochter}, {Prochaska}, \&
  {Burles}}]{prochter06a}
{Prochter}, G.~E., {Prochaska}, J.~X., \& {Burles}, S.~M. 2006, \apj, 639, 766

\bibitem[{{Quider} {et~al.}(2011){Quider}, {Nestor}, {Turnshek}, {Rao},
  {Monier}, {Weyant}, \& {Busche}}]{quider11a}
{Quider}, A.~M., {Nestor}, D.~B., {Turnshek}, D.~A., {et~al.} 2011, \aj, 141,
  137

\bibitem[{{Rubin} {et~al.}(2010){Rubin}, {Prochaska}, {Koo}, {Phillips}, \&
  {Weiner}}]{rubin10a}
{Rubin}, K.~H.~R., {Prochaska}, J.~X., {Koo}, D.~C., {Phillips}, A.~C., \&
  {Weiner}, B.~J. 2010, \apj, 712, 574

\bibitem[{{Sargent} {et~al.}(1988){Sargent}, {Steidel}, \&
  {Boksenberg}}]{sargent88a}
{Sargent}, W.~L.~W., {Steidel}, C.~C., \& {Boksenberg}, A. 1988, \apj, 334, 22

\bibitem[{{Schlegel} {et~al.}(2009){Schlegel}, {White}, \&
  {Eisenstein}}]{schlegel09a}
{Schlegel}, D., {White}, M., \& {Eisenstein}, D. 2009, in ArXiv Astrophysics
  e-prints, Vol. 2010, astro2010: The Astronomy and Astrophysics Decadal
  Survey, 314

\bibitem[{{Schneider} {et~al.}(2010){Schneider}, {Richards}, {Hall}, {Strauss},
  {Anderson}, {Boroson}, {Ross}, {Shen}, {Brandt}, {Fan}, {Inada}, {Jester},
  {Knapp}, {Krawczyk}, {Thakar}, {Vanden Berk}, {Voges}, {Yanny}, {York},
  {Bahcall}, {Bizyaev}, {Blanton}, {Brewington}, {Brinkmann}, {Eisenstein},
  {Frieman}, {Fukugita}, {Gray}, {Gunn}, {Hibon}, {Ivezi{\'c}}, {Kent}, {Kron},
  {Lee}, {Lupton}, {Malanushenko}, {Malanushenko}, {Oravetz}, {Pan}, {Pier},
  {Price}, {Saxe}, {Schlegel}, {Simmons}, {Snedden}, {SubbaRao}, {Szalay}, \&
  {Weinberg}}]{schneider10a}
{Schneider}, D.~P., {Richards}, G.~T., {Hall}, P.~B., {et~al.} 2010, \aj, 139,
  2360

\bibitem[{{Shen} \& {M{\'e}nard}(2012)}]{shen12a}
{Shen}, Y., \& {M{\'e}nard}, B. 2012, \apj, 748, 131

\bibitem[{{Steidel} {et~al.}(1994){Steidel}, {Dickinson}, \&
  {Persson}}]{steidel94a}
{Steidel}, C.~C., {Dickinson}, M., \& {Persson}, S.~E. 1994, \apjl, 437, L75

\bibitem[{{Steidel} \& {Sargent}(1992)}]{steidel92a}
{Steidel}, C.~C., \& {Sargent}, W.~L.~W. 1992, \apjs, 80, 1

\bibitem[{{Tremonti} {et~al.}(2007){Tremonti}, {Moustakas}, \&
  {Diamond-Stanic}}]{tremonti07a}
{Tremonti}, C.~A., {Moustakas}, J., \& {Diamond-Stanic}, A.~M. 2007, \apjl,
  663, L77

\bibitem[{{Tytler} {et~al.}(1987){Tytler}, {Boksenberg}, {Sargent}, {Young}, \&
  {Kunth}}]{tytler87a}
{Tytler}, D., {Boksenberg}, A., {Sargent}, W.~L.~W., {Young}, P., \& {Kunth},
  D. 1987, \apjs, 64, 667

\bibitem[{{Vanden Berk} {et~al.}(2008){Vanden Berk}, {Khare}, {York},
  {Richards}, {Lundgren}, {Alsayyad}, {Kulkarni}, {SubbaRao}, {Schneider},
  {Heckman}, {Anderson}, {Crotts}, {Frieman}, {Stoughton}, {Lauroesch}, {Hall},
  {Meiksin}, {Steffing}, \& {Vanlandingham}}]{vandenberk08a}
{Vanden Berk}, D., {Khare}, P., {York}, D.~G., {et~al.} 2008, \apj, 679, 239

\bibitem[{{Vanden Berk} {et~al.}(2001){Vanden Berk}, {Richards}, {Bauer},
  {Strauss}, {Schneider}, {Heckman}, {York}, {Hall}, {Fan}, {Knapp},
  {Anderson}, {Annis}, {Bahcall}, {Bernardi}, {Briggs}, {Brinkmann}, {Brunner},
  {Burles}, {Carey}, {Castander}, {Connolly}, {Crocker}, {Csabai}, {Doi},
  {Finkbeiner}, {Friedman}, {Frieman}, {Fukugita}, {Gunn}, {Hennessy},
  {Ivezi{\'c}}, {Kent}, {Kunszt}, {Lamb}, {Leger}, {Long}, {Loveday}, {Lupton},
  {Meiksin}, {Merelli}, {Munn}, {Newberg}, {Newcomb}, {Nichol}, {Owen}, {Pier},
  {Pope}, {Rockosi}, {Schlegel}, {Siegmund}, {Smee}, {Snir}, {Stoughton},
  {Stubbs}, {SubbaRao}, {Szalay}, {Szokoly}, {Tremonti}, {Uomoto}, {Waddell},
  {Yanny}, \& {Zheng}}]{vandenberk01a}
{Vanden Berk}, D.~E., {Richards}, G.~T., {Bauer}, A., {et~al.} 2001, \aj, 122,
  549

\bibitem[{{Weiner} {et~al.}(2009){Weiner}, {Coil}, {Prochaska}, {Newman},
  {Cooper}, {Bundy}, {Conselice}, {Dutton}, {Faber}, {Koo}, {Lotz}, {Rieke}, \&
  {Rubin}}]{weiner09a}
{Weiner}, B.~J., {Coil}, A.~L., {Prochaska}, J.~X., {et~al.} 2009, \apj, 692,
  187

\bibitem[{{Weymann} {et~al.}(1979){Weymann}, {Williams}, {Peterson}, \&
  {Turnshek}}]{weymann79a}
{Weymann}, R.~J., {Williams}, R.~E., {Peterson}, B.~M., \& {Turnshek}, D.~A.
  1979, \apj, 234, 33

\bibitem[{{Wild} {et~al.}(2006){Wild}, {Hewett}, \& {Pettini}}]{wild06a}
{Wild}, V., {Hewett}, P.~C., \& {Pettini}, M. 2006, \mnras, 367, 211

\bibitem[{{Yan}(2011)}]{yan11a}
{Yan}, R. 2011, \aj, 142, 153

\bibitem[{{Yip} {et~al.}(2004){Yip}, {Connolly}, {Vanden Berk}, {Ma},
  {Frieman}, {SubbaRao}, {Szalay}, {Richards}, {Hall}, {Schneider}, {Hopkins},
  {Trump}, \& {Brinkmann}}]{yip04a}
{Yip}, C.~W., {Connolly}, A.~J., {Vanden Berk}, D.~E., {et~al.} 2004, \aj, 128,
  2603

\bibitem[{{York} {et~al.}(2000){York}, {Adelman}, {Anderson}, {Anderson},
  {Annis}, {Bahcall}, {Bakken}, {Barkhouser}, {Bastian}, {Berman}, {Boroski},
  {Bracker}, {Briegel}, {Briggs}, {Brinkmann}, {Brunner}, {Burles}, {Carey},
  {Carr}, {Castander}, {Chen}, {Colestock}, {Connolly}, {Crocker}, {Csabai},
  {Czarapata}, {Davis}, {Doi}, {Dombeck}, {Eisenstein}, {Ellman}, {Elms},
  {Evans}, {Fan}, {Federwitz}, {Fiscelli}, {Friedman}, {Frieman}, {Fukugita},
  {Gillespie}, {Gunn}, {Gurbani}, {de Haas}, {Haldeman}, {Harris}, {Hayes},
  {Heckman}, {Hennessy}, {Hindsley}, {Holm}, {Holmgren}, {Huang}, {Hull},
  {Husby}, {Ichikawa}, {Ichikawa}, {Ivezi{\'c}}, {Kent}, {Kim}, {Kinney},
  {Klaene}, {Kleinman}, {Kleinman}, {Knapp}, {Korienek}, {Kron}, {Kunszt},
  {Lamb}, {Lee}, {Leger}, {Limmongkol}, {Lindenmeyer}, {Long}, {Loomis},
  {Loveday}, {Lucinio}, {Lupton}, {MacKinnon}, {Mannery}, {Mantsch}, {Margon},
  {McGehee}, {McKay}, {Meiksin}, {Merelli}, {Monet}, {Munn}, {Narayanan},
  {Nash}, {Neilsen}, {Neswold}, {Newberg}, {Nichol}, {Nicinski}, {Nonino},
  {Okada}, {Okamura}, {Ostriker}, {Owen}, {Pauls}, {Peoples}, {Peterson},
  {Petravick}, {Pier}, {Pope}, {Pordes}, {Prosapio}, {Rechenmacher}, {Quinn},
  {Richards}, {Richmond}, {Rivetta}, {Rockosi}, {Ruthmansdorfer}, {Sandford},
  {Schlegel}, {Schneider}, {Sekiguchi}, {Sergey}, {Shimasaku}, {Siegmund},
  {Smee}, {Smith}, {Snedden}, {Stone}, {Stoughton}, {Strauss}, {Stubbs},
  {SubbaRao}, {Szalay}, {Szapudi}, {Szokoly}, {Thakar}, {Tremonti}, {Tucker},
  {Uomoto}, {Vanden Berk}, {Vogeley}, {Waddell}, {Wang}, {Watanabe},
  {Weinberg}, {Yanny}, \& {Yasuda}}]{york00a}
{York}, D.~G., {Adelman}, J., {Anderson}, Jr., J.~E., {et~al.} 2000, \aj, 120,
  1579

\bibitem[{{York} {et~al.}(2006){York}, {Khare}, {Vanden Berk}, {Kulkarni},
  {Crotts}, {Lauroesch}, {Richards}, {Schneider}, {Welty}, {Alsayyad}, {Kumar},
  {Lundgren}, {Shanidze}, {Smith}, {Vanlandingham}, {Baugher}, {Hall},
  {Jenkins}, {Menard}, {Rao}, {Tumlinson}, {Turnshek}, {Yip}, \&
  {Brinkmann}}]{york06a}
{York}, D.~G., {Khare}, P., {Vanden Berk}, D., {et~al.} 2006, \mnras, 367, 945

\bibitem[{{Zhu} {et~al.}(2009){Zhu}, {Moustakas}, \& {Blanton}}]{zhu09a}
{Zhu}, G., {Moustakas}, J., \& {Blanton}, M.~R. 2009, \apj, 701, 86

\end{thebibliography}
